\documentclass[t, 11pt]{article} 
\usepackage[sectionbib]{natbib}
\usepackage{array,epsfig}
\usepackage[]{hyperref}  
\usepackage{geometry}
\usepackage{amsmath, booktabs, chngcntr}
\usepackage{authblk}
\usepackage{amssymb}
\usepackage{amsfonts}
\usepackage{multirow}
\usepackage{amsthm}
\usepackage{graphicx}
\usepackage{caption,subcaption, array}
\usepackage{float}
\usepackage{bm}
\usepackage{epstopdf, placeins}
\usepackage{color,comment, enumitem}
\newcolumntype{H}{>{\setbox0=\hbox\bgroup}c<{\egroup}@{}}
\usepackage{setspace}
\newtheorem{theorem}{Theorem}
\newtheorem{lemma}{Lemma}

\newtheorem{proposition}{Proposition}
\theoremstyle{definition}

\DeclareMathOperator{\bbeta}{\bm{\beta}}
\DeclareMathOperator{\bX}{\bm{\mathbf{X}}}
\DeclareMathOperator{\bB}{\mathbf{B}}
\DeclareMathOperator{\bA}{\mathbf{A}}
\DeclareMathOperator{\bSigma}{\mathbf{\Sigma}}
\DeclareMathOperator{\bLambda}{\mathbf{\Lambda}}
\DeclareMathOperator{\bkappa}{\bm{\kappa}}

\DeclareMathOperator{\boldeta}{\bm{\eta}}
\DeclareMathOperator{\bdelta}{\bm{\delta}}
\DeclareMathOperator{\bDelta}{\mathbf{\Delta}}

\newcommand{\norm}[1]{\left\Vert #1\right\Vert}
\newcolumntype{H}{>{\setbox0=\hbox\bgroup}c<{\egroup}@{}}

\newcommand{\bred}[1]{{\textcolor{black}{#1}}}

\begin{document}

\title{\bf Sparse Sliced Inverse Regression \\ via Cholesky Matrix Penalization}
\author[1,3]{Linh H. Nghiem}
\author[1]{Francis K.C.Hui}
\author[2,3]{Samuel M{\"u}ller}
\author[1]{A.H. Welsh}

\affil[1]{Research School of Finance, Actuarial Studies and Statistics, Australian National University}
\affil[2]{Department of Mathematics and Statistics, Macquarie University}
\affil[3]{School of Mathematics and Statistics, University of Sydney}
\date{}
\maketitle

\begin{quotation}
\noindent {\it Abstract:}
We introduce a new sparse sliced inverse regression estimator called Cholesky matrix penalization and its adaptive version for achieving sparsity in estimating the dimensions of the central subspace. The new estimators use the Cholesky decomposition of the covariance matrix of the covariates and include a regularization term in the objective function to achieve sparsity in a computationally efficient manner. We establish the theoretical values of the tuning parameters that achieve estimation  and variable selection consistency for the central subspace. Furthermore, we propose a new projection information criterion to select the tuning parameter for our proposed estimators and prove that the new criterion facilitates selection consistency. The Cholesky matrix penalization estimator inherits the strength of the Matrix Lasso and the Lasso sliced inverse regression estimator; it has superior performance in numerical studies and can be adapted to other sufficient dimension methods in the literature.

\vspace{9pt}

\par
\end{quotation}\par

\def\thefigure{\arabic{figure}}
\def\thetable{\arabic{table}}

\renewcommand{\theequation}{\thesection.\arabic{equation}}

\fontsize{12}{14pt plus.8pt minus .6pt}\selectfont

\section{Introduction}

In a regression problem with a scalar outcome $y$ and a $p$-variate predictor ${\bX} = (X_1, \ldots, X_p)^\top$, sufficient dimension reduction refers to a class of methods that try to express the outcome as a function of a few linear combinations of covariates \citep{li2018sufficient}. In other words, sufficient dimension reduction aims to find a matrix  ${\bB}$ of dimension $p \times d$ with $d<<p$ such that
\begin{equation} 
y \perp {\bX} \mid {\bB}^\top {\bX},
\label{eq:sufficient dimension reduction}
\end{equation}
with $\perp$ denoting statistical independence. Condition \eqref{eq:sufficient dimension reduction} implies that the $d$ linear combinations ${\bB^\top \bX}$ contain all the information about $y$ on ${\bX}$, so we can replace ${\bX}$ by ${\bB^\top \bX}$ without loss of information. Dimension reduction is achieved because the number of linear combinations $d$ is usually much smaller than the number of covariates $p$.   Let $\bbeta_1, \ldots, \bbeta_d$ be the columns of ${\bB}$, an alternative formulation of the relationship between $y$ and $\bX$ under \eqref{eq:sufficient dimension reduction} is provided by the multiple index model 
\begin{equation} 
y = f(\bbeta_1^\top {\bX}, \ldots, \bbeta_d^\top {\bX}, \varepsilon),
\label{eq:MIM}
\end{equation}
where $f$ is an unknown link function and $\varepsilon$ is a random noise term independent of $ {\bX}$. In  \eqref{eq:sufficient dimension reduction} and \eqref{eq:MIM}, the matrix  $\bB$ and the vectors $\bbeta_1, \ldots, \bbeta_d$ are generally not unique when $d\geq1$ \citep{li2018sufficient}.
Therefore, the goal in sufficient dimension reduction is to identify the central subspace, which is defined as the intersection of all subspaces spanned by column spaces of $ {\bB}$ satisfying \eqref{eq:sufficient dimension reduction}. The central subspace, denoted by $\mathcal{S}_{y\perp  {\bX}}$, is unique under mild conditions \citep{li2018sufficient}. The transformations $ \bbeta_j^\top {\bX}, ~j=1,\ldots, d$, are called sufficient predictors. The number of indices in the multiple index model, $d$, is also known as the structural dimension of the central subspace. A variety of sufficient dimension reduction methods have been proposed in the literature, including sliced inverse regression (SIR, \citet{li1991sliced}),  sliced average variance estimation  (SAVE, \citet{dennis2000save}), principal Hessian direction (pHd, \citep{li1992principal,cook1998principal}), minimum average variance estimation \citep{xia2002adaptive}, and directional regression  \citep{li2007directional}, among others. An overview of these methods can be found in \citet{li2018sufficient}. 

When the number of covariates is large, we often assume each dimension $ \bbeta_1, \ldots,  \bbeta_d$ is sparse, that is, for each $j=1,\ldots,d$, only a few elements of each dimension $ {\bbeta}_j = (\beta_{j1},\ldots, \beta_{jp})^\top$ are non-zero. Motivated by this idea, the last few years have seen an emerging literature combining sparsity with SIR as well as with other sufficient dimension reduction methods through adding a regularization term to an appropriate objective function. For example, \cite{yin2015sequential} proposed a sequential approach for estimating SIR. \cite{lin2018consistency} first proposed a screening approach to perform variable selection; then, the selected variables were included in the classical SIR. Assuming the covariates follow the standard $p$-variate Gaussian distribution with at most $s$ non-zero components in each dimension, \cite{lin2017optimality} established the minimax rate of the risk of the estimated projection matrix  when the number of indices $d$ is bounded. \cite{tan2018convex} proposed a convex formulation for sparse SIR in the high-dimensional setting by adapting techniques from sparse canonical correlation analysis. While most methods have been able to identify the theoretical values of the regularization parameter to obtain estimation consistency of the central space, only a few papers have been able to establish the theoretical values of regularization parameters for variable selection consistency in each estimated dimension; one example is \cite{qian2019sparse}. Even in that case, the method for selecting the regularization parameter in their numerical studies and data application did not guarantee  selection consistency.  

\bred{In this paper, we propose a new approach to constructing a sparse SIR estimator based on the Cholesky decomposition of the sample covariance matrix. This Cholesky matrix penalization (CHOMP) estimator has a close connection to the Lasso SIR estimator proposed by \citet{Linsparse}, but has several advantages over it. First, while both  CHOMP and Lasso SIR achieve estimation consistency for the central subspace, we generalize the CHOMP estimator to an adaptive version that can achieve both estimation and variable selection consistency. Furthermore, for both CHOMP and its adaptive version, we propose a new projection information criterion (PIC) to select the regularization parameter in the corresponding objective function; as far as we are aware, this is the first data-driven method that is theoretically demonstrated to achieve variable selection consistency for the central subspace. Our simulation studies show that the adaptive CHOMP estimator with the regularization parameter selected by the PIC has superior performance to the Lasso SIR in both estimation error and variable selection. Finally, the CHOMP-type estimator is easily generalized to many other sufficient dimension reduction methods, such as SAVE and pHd, and the corresponding adaptive CHOMP-type estimators are shown empirically to have competitive performance in finite samples as well.}

The following notation is used throughout the paper. For any $p$-dimensional non-zero vector $ \bm{v} = (v_1,\ldots, v_p)^\top$, let $\mathcal{P}( \bm{v}) =  \bm{v}( \bm{v}^\top  \bm{v})^{-1} \bm{v}^\top$ denote the projection matrix associated with $ \bm{v}$, and let $\norm{ \bm{v}}_2 = \left(\sum_{j=1}^p v_j^2\right)^{1/2}$, $\norm{ \bm{v}}_1 = \sum_{j=1}^{p} \vert v_j \vert$, $\norm{\bm {v}}_0 = \sum_{j=1}^{p} 1(v_{j}\neq 0)$, and $\norm{ \bm{v}}_\infty = \max_j \vert v_j\vert$ denote its $\ell_2$, $\ell_1$, $\ell_0$ and $\ell_\infty$ norm respectively. For any index set $T$, the notation $ \bm{v}_T$ and $ \bm{v}_{T^c}$ denote the sub-vectors consisting of the components of $ \bm{v}$ in $T$ and $T^c$, respectively. For any $m \times q$  non-singular matrix  $ {\bA}$ with entries $a_{ij}$, let $\mathcal{P}( {\bA}) =  {\bA(\bA^\top \bA)}^{-1} {\bA}^\top$ denote the projection matrix  associated with $ {\bA}$. Also, we define the Frobenius norm of $ {\bA}$ to be $\norm{ {\bA}}_F =  \left(\sum_{i=1}^m\sum_{j=1}^q a_{ij}^2\right)^{1/2}$, while the $\ell_2$ induced norm $\norm{ {\bA}}_2$ is its largest singular value $\sigma_1( {\bA})$. Finally, for ease of notation, we let $\mu_j$ generically denote the tuning parameter used to estimate the $j^{th}$ dimension of the central subspace for all the penalized methods.

\section{A review of sliced inverse regression and the Matrix Lasso}
\label{review}
We first review sliced inverse regression (SIR), which is the basis for other methods discussed in this paper. Assuming the predictor vector $ \mathbf{X}$ follows an elliptical distribution with location vector zero and scale matrix  $ {\bSigma}$, it was demonstrated in \cite{li1991sliced} that the column space of $ \mathbf{B}$ in equation \eqref{eq:sufficient dimension reduction}  satisfies 
\begin{equation}
\bSigma \text{col}({\bB}) = \text{col}( \mathbf{\Lambda}),
\label{eq:sliced inverse regression}
\end{equation}	
where $ \bLambda = \text{var}\left\{\mathbb{E}( {\bX}|y)\right\}$. If we observe independent and identically distributed data pairs $( \bm{x}_i^\top,y_i), ~i=1,\ldots,n$ with $ \bm{x}_i = (x_{i1},\ldots, x_{ip})^\top$, let $\mathcal{X}$ denote the $n \times p$ design matrix; without loss of generality, assume each predictor is centered at zero and let $\hat{ {\bSigma}} = n^{-1}\sum_{i=1}^{n}  \bm{x}_i  \bm{x}_i^\top = n^{-1} \mathcal{X}^\top \mathcal{X}$ be the sample covariance matrix. Next, if the outcome $y$ is ordered, then the matrix  $ {\bLambda}$ is estimated by first dividing the data into $H$ non-overlapping slices of roughly equal sizes, $J_1, \ldots, J_H$, based on the increasing order of $y$. If $y$ is categorical, each slice may correspond to one category in the outcome. Then, we compute the vector of covariate averages within each slice, $\bar{ \bm{x}}_h^\top = \vert J_h \vert^{-1} \sum_{i=1}^{n}  \bm{x}_i^\top1(y_i \in J_h)$, with $\vert J_h \vert$ being the size of the slice $J_h$. As a result, an estimate for $ {\bLambda}$ is given by
$\hat{ \bLambda}=H^{-1}\sum_{h=1}^{H}  \bar{\bm{x}}_h\bar{ \bm{x}}_h^\top. 
$ 
Let $\hat{ \boldeta}_1, \ldots, \hat{ \boldeta}_d$ be the eigenvectors corresponding to the $d$ largest eigenvalues  of $\hat{ \bLambda}$. Then \cite{li1991sliced} showed that each dimension $\hat{ \bbeta}_j$ of the central subspace can be estimated by
\begin{equation} 
\hat{ \bbeta}_j = \hat{\bSigma}^{-1} \hat{ \boldeta}_j, ~j= 1,\ldots,d.
\label{eq:nosparse}
\end{equation}

In a recent paper, \cite{Linsparse} introduced two sparse SIR estimators that are closely connected to the Lasso estimator in the regular linear model, namely the Matrix  Lasso and the Lasso sliced inverse regression (Lasso SIR). Based on the relationship \eqref{eq:nosparse}, the Matrix  Lasso estimator $\hat{ \bbeta}^{\text{ML}}_j$ is defined as 
\begin{equation}
\hat{ \bbeta}^{\text{ML}}_j = \arg\min_{ \bbeta_j} \dfrac{1}{2} \norm{\hat{ {\boldeta}}_j-\hat{ {\bSigma}} \bbeta_j}_2^2+\mu_j\norm{ \bbeta_j}_1,~j=1,\ldots,d.
\label{eq:matrix lasso}
\end{equation}
Although the Matrix Lasso estimator was introduced in \cite{Linsparse}, it was largely dismissed and its theoretical properties are yet to be examined. One possible reason for this was because, similar to any regularization method, the performance of matrix  Lasso depends on how the tuning parameters $\mu_j,~ j=1,\ldots, d,$ are chosen. However, selecting appropriate tuning parameters for the Matrix  Lasso is challenging from both theoretical and practical perspectives for two reasons.  First, the outcome $ \hat{\boldeta}_j$ does not contain independent observations, so regular cross-validation is not guaranteed to work. Also, unlike the linear model case, the matrix  $\hat{ {\bSigma}}$ is a $p \times p $ symmetric matrix, so the first term in \eqref{eq:matrix lasso} can be zero if $\hat{ {\bSigma}}$ is invertible as occurs for example in low-dimensional settings where $n>p$. 

On the other hand, a main advantage of the Matrix  Lasso that was not emphasized in \cite{Linsparse} is that the formulation \eqref{eq:matrix lasso} is convex and directly mimics the population relationship that characterizes SIR in \eqref{eq:sliced inverse regression}. As a result, the formulation of the Matrix  Lasso can be adapted to many other sufficient dimension reduction methods that are formed by changing the matrix  $ {\bLambda} = \text{var}\left\{\mathbb{E}( {\bX}|y)\right\} $ in equation \eqref{eq:sliced inverse regression} to another quantity. For example, a sliced average variance estimator is obtained with $ {\bLambda}=  \mathbb{E}\left\{{\bSigma}-\text{var}( {\bX}|y)\right\}^2$, or a principal Hessian direction estimator is obtained with $ {\bLambda}= \mathbb{E}\left[ {\bX\bX^\top} \left\{y-\mathbb{E}(y)\right\}\right]$. Hence, understanding the behavior of the Matrix  Lasso estimator and building upon it to devise improved estimators provides a unified strategy for investigating sparse dimensions of the central subspace, as will be shown in Section \ref{sir-otherir}.

In fact, the Lasso SIR estimator, also proposed by \cite{Linsparse}, can be considered as a recasting  of the Matrix  Lasso. Essentially, using a special formulation of the matrix $\hat{\bLambda}$ as an estimate of $\bLambda = \text{var}\{\mathbb{E}(\bX \vert y)\}$, \cite{Linsparse} proves that each eigenvector can be computed as $\hat{\boldeta}_j = n^{-1}\mathcal{X}^\top\tilde{\bm{y}}_j$ for an appropriately defined vector $\tilde{\bm{y}}_j, ~j = 1,\ldots, d$; see section C of the Supplementary Material for more details. If we use the sample covariance matrix  $\hat{\bSigma} = n^{-1}\mathcal{X}^\top\mathcal{X}$ to estimate $\bSigma$, then \eqref{eq:sliced inverse regression} can be written as $\mathcal{X}^\top \mathcal{X} {\bbeta}_j \propto  \mathcal{X}^\top \mathcal{X} \tilde{\bm {y}_j}$ and the Lasso SIR estimator is defined as
\begin{equation*}
\hat{ {\bbeta}}^{\text{L}}_j=\arg\min_{ {\bbeta}}\frac{1}{2 n}\left\|\widetilde{\bm{y}_j}-\mathcal{X} {\bbeta}\right\|_{2}^{2}+\mu_j\|{\bbeta}\|_{1}, ~ j=1,\ldots, d.
\label{eq:Lassosir}
\end{equation*} 
This formulation depends on the special form of $\bLambda = \text{var}\{\mathbb{E}(\bX \vert y)\}$ used in SIR. Therefore, it is not straightforward how it can be adapted to other sufficient dimension reduction methods that are obtained by changing $\bLambda$. In the next section, we provide another reformulation of the Matrix  Lasso that both inherits desirable properties of the Lasso SIR and can be applied to other methods in a more straightforward way. 

\section{Cholesky matrix penalization for sliced inverse regression} \label{methods}
\subsection{Estimators}
Recall that at the population level, the SIR estimator satisfies the relationship \eqref{eq:sliced inverse regression}. Let $ {\boldeta}_1, \ldots,  {\boldeta}_d$ be the eigenvectors associated with the  $d$ largest eigenvalues of $ {\bLambda}$, then the vector $ {\bbeta}_j$ satisfies
\begin{equation}
{\bSigma} \bbeta_j =  \boldeta_j, ~ j=1,\ldots, d.
\label{eq:sliced inverse regression2}
\end{equation}
For each $j$, equation \eqref{eq:sliced inverse regression2} is a system of $p$ linear equations. Because we do not impose any additional structure on the symmetric and positive definite matrix $\bSigma$, an efficient way to solve the system is through the Cholesky decomposition. Specifically, letting $ {\bSigma} =  \bm{L} \bm{L}^\top$, where $ \bm{L}$ is the  Cholesky factor of $ {\bSigma}$, then equation \eqref{eq:sliced inverse regression2} is  equivalent to
\begin{equation*}
\bm{L}^\top  {\bbeta}_j =  \bm\kappa_j, ~ \text{where ~}  \bm{L} \bkappa_j =  {\boldeta}_j,~ j=1,\ldots, d.
\label{eq:Cholesky1} 
\end{equation*}
Since $ \bm{L}$ is a lower triangular matrix, the vector $ \bkappa_j$ is obtained by backward substitution, and the vector $ \bbeta_j$ is obtained  by  forward substitution. Next, denote $\hat{\bm{L}}$ and $\hat{\boldeta}_j$ as estimators for $\bm{L}$ and the eigenvector $ {\boldeta}_j$ respectively. 
Typically, the vector $\hat{  \boldeta}_j$ is the eigenvector of the matrix  $\hat{ {\bLambda}}$. 

Let $\hat{ \bkappa}_j$ be calculated from $ \hat{ \bm{L}} \hat{ \bkappa}_j = \hat{ {\boldeta}}_j$; for $\hat\bkappa_j$ to be well-defined, the estimator $\hat{\bm{L}}$ needs to be invertible. For the remaining of the paper, we assume $n>p$, so we can choose $\hat{\bm{L}}$ to be the Cholesky factor of the sample covariance matrix $\hat{\bSigma}$. In Section D of the Supplementary Material, we investigate a high dimensional setting $(n<p)$ where the Cholesky factor $\bm{L}$ can be efficiently estimated by imposing an additional structure on $\bSigma$. We define the Cholesky matrix penalization (CHOMP) estimator for the SIR to be
\begin{equation}
\hat{ \bbeta}_j =  \arg\min_{ \bbeta_j} \dfrac{1}{2} \norm{\hat{ \bm{L}}^\top {\bbeta}_j - \hat{ {\bkappa}}_j}_2^2 + \mu_j \norm{ {\bbeta}_j}_1, ~ j=1,\ldots,d, 
\label{eq:Cholesky matrix  penalization}
\end{equation}
where $\mu_j$ is a non-negative tuning parameter. Furthermore, we can penalize each component of $
\bbeta_j$ differently by introducing a vector of adaptive weights $ \omega_j = (\omega_{j1}, \ldots, \omega_{jp})^\top$ and defining
\begin{equation}
\hat{ \bbeta}_j^* = \arg\min_{ \bbeta_j} \dfrac{1}{2} \norm{\hat{ \bm{L}}^\top {\bbeta}_j - \hat{ {\bkappa}}_j}_2^2 + \mu_j \sum_{k=1}^{p} \omega_{jk} \vert \beta_{jk} \vert , ~ j=1,\ldots,d. 
\label{eq:adaCholesky matrix  penalization}
\end{equation}
We refer to this estimator as the adaptive Cholesky matrix  penalization (adaptive CHOMP) estimator, in line with the adaptive Lasso estimator proposed by \cite{zou2006adaptive}. Moreover, similar to \citet{zou2006adaptive}, we set the weights to be $\omega_{jk} = \vert \bar{\beta}_{jk} \vert^{-\gamma}$, with $\bar{\beta}_{jk}$ being the $k${th} component of an initial consistent estimate $\bar{ {\bbeta}}_j$ and $\gamma$ a positive constant. Because $n>p$, we  choose $\bar{ {\bbeta}}_j$ to be the unpenalized estimate $\bar{ {\bbeta}}_j = \hat{ \bSigma}^{-1}\hat{ \boldeta}_j$. \bred{In the simulation study presented in Section \ref{sim}, we find that, as expected, the inclusion of these adaptive weights makes performance of the adaptive CHOMP superior to that of (the unweighted) CHOMP, the Matrix Lasso and the Lasso SIR estimator in terms of both estimation error and variable selection for the central subspace. }


\subsection{Matrix Lasso, Cholesky matrix penalization, and Lasso sliced inverse regression}
The Matrix Lasso, CHOMP and the Lasso SIR estimators essentially derive from the same relationship \eqref{eq:sliced inverse regression}. Moreover, if no regulation is imposed, $\mu_j = 0$, all the estimators are equivalent. However, when regularization is needed to achieve sparse solutions, the behavior of the tuning parameters for the Matrix Lasso are fundamentally different from the other two. 

In fact, from the definition of the Matrix Lasso estimator given in equation \eqref{eq:matrix lasso} and the first-order optimality condition, each component of the Matrix Lasso estimator $\hat{ \bbeta}_{j}^{\text{ML}}=(\hat{\beta}_{j1}^{\text{ML}},\ldots, \hat{\beta}_{jp}^{\text{ML}})^\top$ satisfies
\begin{equation}
\hat{ \bSigma}_k^\top \left(\hat{ \bSigma} \hat{ \bbeta}_{j}^{\text{ML}} - \hat{ \boldeta}_{j}\right) + \mu_j b_{jk}^{\text{ML}} = 0, \quad k=1,\ldots, p,
\label{kkt2}
\end{equation} 
where $\hat{ \bSigma}_k^\top$ denotes the $k$th row of $\hat{ \bSigma}$, the scalar $b_{jk}^{\text{ML}}= \text{sign}(\hat\beta_{jk}^{\text{ML}})$ if $\hat{\beta}_{jk}^{\text{ML}} \neq 0 $ and $b_{jk}^{\text{ML}} \in [-1,1]$ if $\hat{\beta}_{jk}^{\text{ML}} = 0$. As a result, the entire vector $\hat\bbeta_{j}^{\text{ML}}$ is set to  zero if and only if $\mu_j > \norm{\hat{ \bSigma}^\top \hat{ \boldeta}_{j}}_\infty $. Note that each component $\hat{ \bSigma}_k^\top \hat{ \boldeta}_{j}$ is the sum of $p$ terms, so it can grow to infinity when the dimension $p$ grows. This fact does not change when each covariate is standardized to have variance one and the sample covariance matrix is a correlation matrix. As a result, when $p$ is growing, the range of $\mu_j$ that needs to be considered is unbounded. Even in the special case where $\hat{\bSigma}$ is diagonal with elements $\hat{\sigma}_1^2, \ldots, \hat{\sigma}_p^2$, equation \eqref{kkt2} reduces to $\hat\sigma_k^4 \hat{\beta}_{jk}^{\text{ML}} = \hat\sigma_k^2\hat\eta_{jk} - \mu_j b_{jk}^{\text{ML}}$, so $\hat{\beta}_{jk}^{\text{ML}} = 0 $ when $\mu_j \geq  \hat\sigma_k^2\hat\eta_{jk}$; in other words, the range of $\mu_j$ to be considered is affected by the scale of the covariates.

On the other hand, the range of $\mu_j$ for both the CHOMP and the Lasso SIR estimator that needs to be considered is the unit interval. Each component of the CHOMP estimate $\hat{\bbeta}_j$ and of the Lasso SIR estimate $\hat{\bbeta}_j^{L}$ satisfies
\begin{align}
\hat{\bm{L}}_k^\top \left(\hat{\bm{L}}\hat{ \bbeta}_{j} - \hat{ \bkappa}_{j}\right) + \mu_j b_{jk} = 0, \quad \text{i.e} \quad \hat{\bSigma}_k^\top \hat{\bbeta}_j -  {\eta}_{jk}  + \mu_j b_{jk} = 0
\label{kkt} \\
n^{-1} \bm{x}_k^\top \left(\mathcal{X}\hat{ \bbeta}_{j}^{L} - \tilde{y}_{j}\right) + \mu_j b_{jk}^L = 0 \quad \text{i.e} \quad \hat{\bSigma}_k^\top \hat{\bbeta}_j^{L} -  {\eta}_{jk}  + \mu_j b_{jk}^{L} = 0
\label{kkt3} 
\end{align}
correspondingly, where $\hat{\bm{L}}_k^\top$ denotes the $k$th row of $\hat{\bm{L}}$, the scalar $b_{jk} = \text{sign}(\hat\beta_{jk})$ if $\hat\beta_{jk} \neq 0 $ and $b_{jk} \in [-1,1]$ if $\hat{\beta}_{jk} = 0$, with similar definition for $b_{jk}^L$. This implies that the CHOMP and the Lasso SIR estimators have the same estimating equation for every tuning parameter $\mu_j$.  As a result, the whole vector $\hat{\bbeta}_j = 0$  is set to zero if and only if $ \mu_j \geq  \norm{\hat{ \bm{L}} \hat{ \bm{\kappa}}_j}_\infty =  \norm{\hat{\boldeta}_{j}}_\infty $. Since all the components of $\hat{ {\boldeta}}_j$ are between $-1$ and $1$, to choose the appropriate value for $\mu_j$, we only need to consider $\mu_j \in [0, 1]$, regardless of the dimension $p$. In practice, we usually choose the tuning parameter from a grid of values, so having a fixed upper bound on the grid regardless of $p$ is desirable to fine tune the estimator. In the special case where the matrix $\hat{\bSigma}$ is diagonal, both equations \eqref{kkt} and \eqref{kkt3} become $\hat\sigma_k^2 \hat\beta_{jk} = \hat\eta_{jk} - \mu_j b_{jk}$. In this case, the component $\hat{\beta}_{jk}$ is set to zero if $\mu_j > \hat{\eta}_{jk}$; hence, variable selection is  done by thresholding the magnitude of the component of the eigenvector $\hat{\boldeta}_j$.  

One way to restrict the bound of the tuning parameters for the Matrix Lasso estimator is to work with the standardized covariates $\bm{z}_i = \hat{ {\bSigma}}^{-1/2}  \bm{x}_i, ~i=1,\ldots, n$. In that case, since the sample covariance matrix of the transformed $z$-data is the identity matrix, the quantity $\norm{\hat{\bSigma}\hat{\boldeta}_j}_\infty = \hat{\boldeta}_j$ is bounded by one. However, a major disadvantage of this approach is that the Matrix Lasso estimator on the $\bm{z}_i$-data, denoted as $\hat\bbeta_j^{\text{ML}(z)}$, can be sparse, but the final estimator $\hat{\bbeta}_j^{\text{ML}} = \hat\bSigma^{-1/2} \hat\bbeta_j^{\text{ML}(z)}$ for $ {\bbeta}_j$ is not guaranteed to be sparse because we do not impose any sparsity requirement on the matrix $\hat{ {\bSigma}}^{-1/2}$. As a result, no variable selection is achieved for any dimension of the central subspace.   

Finally, unlike the Lasso SIR estimator, the CHOMP estimator inherits the flexibility of the Matrix Lasso in that it is easy to adapt to other sufficient dimension reduction methods. For example, for sliced average variance estimator, we change $\bLambda = \text{var} \{E (\bX | y) \}\}$ in equation \eqref{eq:sliced inverse regression} to $\bLambda = \mathbb{E}\left\{\bSigma - \text{var}(\bX \vert y)\right\}^2$ and make the corresponding estimate $\hat\bLambda$ its sample version. In this situation, it is not as straightforward to define the vector $\tilde{\bm{y}}_j$ such that the eigenvector $\hat{\boldeta}_j$ of $\hat{\bLambda}$ can be written as $\hat\boldeta_j = n^{-1} \mathcal{X}^\top \tilde{\bm{y}}_j$ to apply the idea of the Lasso SIR. Nevertheless, we can still compute the CHOMP estimate and its adaptive version by solving the problem \eqref{eq:Cholesky matrix penalization}. We will elaborate more in Section \ref{sir-otherir}. 

\subsection{Projection information criterion}
\label{tuning}

To choose the tuning parameter $\mu_j$ for the CHOMP and adaptive CHOMP estimators, we propose to minimize the  projection information criterion (PIC) defined as
\begin{equation}
\textsc{PIC}(\mu_j) = \begin{cases}
\norm{\mathcal{P}\left\{\hat{ \bbeta}_j({\mu_j})\right\}-\mathcal{P}(\bar{ \bbeta}_j)}_F^2 + \dfrac{\log(p)}{p} \norm{\hat{ \bbeta}_j({\mu_j})}_0,& ~ \text{if}~ \hat{ \bbeta}({\mu_j}) \neq  \bm{0}  \label{first} \\
\infty, & ~ \text{if} ~ \hat{ \bbeta}_j({\mu_j}) =  \bm{0}, 
\end{cases}
\end{equation}
where for the $j$th dimension, the notation $\hat{\bbeta}_j({\mu_j})$ denotes either the CHOMP or its adaptive version associated with the tuning parameter $\mu_j > 0$. The main difference between the projection and the usual information criteria is in the loss function, and we motivate our choice as follows. In the multiple index model, each vector $ {\bbeta}_j$ is not unique, but the projection matrix associated with it is unique. Hence, a sensible way of quantifying the goodness of fit is via the estimated projection matrix. The specific form of the loss part, $\norm{\mathcal{P}\left\{\hat{ \bbeta}_j({\mu_j}) \right\}-\mathcal{P}(\bar{ \bbeta}_j)}_F^2$, measures the deviation of $\hat{ \bbeta}({\mu_j})$ from an already-established consistent estimator. Because the projection matrix is not well-defined for the zero vector, we ignore this case by setting the PIC to infinity when parameter estimates are zero. In other words, we do not expect the true vector $ \bbeta_j$ to be a zero vector for any dimension. The model complexity penalty term $\tau_j = \log(p)/p$ controls the trade-off between model loss and complexity part. \bred{This choice of model complexity penalty has the following intuition. Because the loss part is bounded above by $2$ and the number of non-zero components for each $\hat{\bbeta}_j(\mu_j)$ can range from zero to $p$, the denominator of $\tau_j$ is set to $p$ to make the two parts have relatively the same magnitude. The numerator of $\tau_j$ follows the same spirit as the Bayesian information criterion (BIC) penalty; however, it is set to $\log(p)$ instead of to $\log(n)$ to make $\tau_j$ go to zero without imposing any further condition on the growth rate of $n$ and $p$.} 
For each dimension $j=1,\ldots, d$, we demonstrate in Section \ref{theory} that this model complexity term leads to selection consistency; i.e. PIC asymptotically identifies the non-zero components of each dimension $ {\bbeta}_j$ correctly with this model complexity term.

Finally, we briefly mention the issue of estimating the number of indices $d$ from the data. In general, if the original data is divided into $H$ slices, the maximum number of dimensions that can be estimated by  the SIR methods is $H-1$ \citep{li2018sufficient}. When $p$ is fixed, a variety of methods for determining $d$ are proposed in the literature, including the sequential testing approach of \cite{li1991sliced} and the bootstrap methods of \cite{ye2003using}, among others. When $p$ is growing , \cite{Linsparse} proposed a method for choosing the number of indices for the Lasso SIR based on a clustering of the eigenvalues of $\hat{ {\bLambda}}$. We anticipate similar methods could be developed for (adaptive) CHOMP estimators, but the precise choice of $d$ is outside the scope of the paper, and in the numerical studies below, we assume $d$ to be known. 

\section{Theoretical Results} \label{theory}
We prove several results related to the estimation consistency and variable selection consistency of the estimated projection matrix  $\mathcal{P}(\hat{ {\bB}})$ and $\mathcal{P}(\hat{ {\bB}}^*)$, where  $ \hat{ {\bB}}$ and $\hat{ {\bB}}^*$ are $p \times d$ matrices whose columns are CHOMP and adaptive CHOMP estimators respectively. These theoretical results are derived by combining results for the Lasso estimator for the regular linear model with the results of the Lasso SIR estimator developed in \cite{Linsparse}. 
Furthermore, we demonstrate that using the new PIC leads to a selection consistent estimator. In this section, we allow the number of covariates $p$ to grow with the sample size $n$, but the ratio $p/n \rightarrow 0$ when $n \rightarrow \infty$. This condition ensures that the Cholesky factor $\hat{\bm{L}}$ of the sample covariance matrix $\hat{\bSigma}$ is invertible with probability one, so the vector $\hat{\bm\kappa}_j$ (and functions thereof) is well-defined. The proofs of all the results can be found in Section A and B of the Supplementary Material.

First, we state  technical conditions that are used throughout the development below. 
\begin{enumerate}[label=(C\arabic*)]	
	\item There exist constants $C_\text{min}$ and $C_\text{max}$ such that $ 0 < C_\text{min} < \lambda_\text{min}( \bSigma) < \lambda_\text{max}( \bSigma) < C_\text{max} < \infty $, where $\lambda_\text{min}( \bSigma)$ and $\lambda_\text{max}( \bSigma)$ denote the minimum and maximum eigenvalue of $ {\bSigma}$, respectively.
	\label{assumption:A1}
	
	\item The $d$ largest eigenvalues $\lambda_1, \ldots, \lambda_d$  of $ {\bLambda}=\text{var}\{\mathbb{E}( {X}|y)\}$ satisfy $0 < \lambda_d \leq \ldots \leq \lambda_1 \leq \lambda_\text{max}( \bSigma) < \infty $. 
	\label{assumption:A2}
	
	\item	The central curve $ {m}(y) = \mathbb{E}( {\bX} \mid y)$ satisfies the sliced stability condition of \cite{lin2018consistency}.	
	\label{assumption:A3}
\end{enumerate}
Condition \ref{assumption:A1} is usually imposed in the analyses of high dimensional linear regression models \citep{wainwright2019high}. This condition implies that the sample covariance matrix $\hat{ {\bSigma}}$ satisfies a so-called restricted value condition over a cone set, which is described more clearly below. As discussed in \cite{Linsparse}, Condition \ref{assumption:A2} is a refined version of a commonly imposed condition in the SIR literature, that is, $\text{rank}{( {\bLambda})}=d$, meaning the dimension of the space spanned by the central curve equals the dimension of the central subspace. Finally, Condition \ref{assumption:A3} controls the smoothness of the central curve $ {m}$ and the tail distribution of $ {m}(y)$; see \cite{lin2018consistency} for a detailed discussion of this condition.

Recall for each dimension $j=1,\ldots, d$, the vector $ {\boldeta}_j =  {\bSigma} \bbeta_j =  {\bm{L}\bm{L}^\top} \bbeta_j$ is the eigenvector associated with the $j$th largest eigenvalue of $ {\bLambda} = \text{var}\{\mathbb{E}( \bX \mid y)\}$, while $\hat{\boldeta}_j$ is the same quantity of the estimated matrix $\hat{ {\bLambda}}$. Define $ \tilde{ {\boldeta}}_j= \mathcal{P}( {\boldeta}_j)\hat{ {\boldeta}}_j$ to be the projection of $\hat{ {\boldeta}}_j$ on the $ {\boldeta}_j$. The projection implies that $\tilde{ {\boldeta}}_j \propto  {\boldeta}_j$.  As a result, if we define $ \tilde{ \bbeta}_j =  {\bSigma}^{-1}\tilde{ {\boldeta}}_j$, then
$
\tilde{ \bbeta}_j  =  {\bSigma}^{-1} \mathcal{P}( {\boldeta}_j)\hat{ {\boldeta}}_j = {\bSigma}^{-1}  {\boldeta}_j( {\boldeta}_j^\top  {\boldeta}_j)^{-1} {\boldeta}_j^\top \hat{ {\boldeta}}_j \propto  \bbeta_j
$; in other words, $\tilde{ \bbeta}_j$ has the same projection matrix as the true dimension $ {\bbeta}_j$.  We refer to $\tilde{ \bbeta}_j$ as the ``pseudo-true'' parameter for the dimension $j$
in the theoretical development and bound the difference $ {\delta}_j = \hat{ \bbeta}_j - \tilde{ {\bbeta}}_j$ to establish the consistency of the estimated projection matrix. 

Denote $S_j=\{k: \beta_{jk}\neq 0\}$, the set of indices corresponding to non-zero components of the true dimension $\bbeta_j$, and $s_j = |S_j|$, the cardinality of the set $S_j$. Furthermore, denote $S = \bigcup_{j=1}^{d} S_j$, the set of active covariates across all dimensions, and $s = \vert S\vert$.  Because $\tilde{ \bbeta}_j \propto  {\bbeta}_j$, then $\tilde{\beta}_{jk}=0$ for any $j \in S_j^c$.  The following theorem establishes the consistency of the estimated projection matrix from the CHOMP estimator.
\begin{theorem}
	Consider a multiple index model with $n\lambda_d = p^\nu$ for $\nu > 1$. Assume Conditions \ref{assumption:A1}-\ref{assumption:A3} hold and the number of dimensions $d$ is known. Let $\hat{ {\bB}}$ be the matrix whose columns $\hat{ \bbeta}_1,  \ldots, \hat{ \bbeta}_d$ are solutions of \eqref{eq:Cholesky matrix penalization} with tuning parameter $\mu_j = M\left\{{\log(p)}/(n\hat{\lambda}_j)\right\}^{1/2}$ for a sufficiently large constant $M$, where $\hat{\lambda}_j$ is the $j$th largest eigenvalue of the matrix $\hat{\bLambda}$. Then,    
	the estimated projection matrix $\mathcal{P}(\widehat{ {\bB}})$  satisfies 
	\begin{equation*}
	\left\|\mathcal{P}(\widehat{ {\bB}})-\mathcal{P}( {\bB}) \right\|\leq C \left\{\frac{s \log (p)}{n \lambda_d}\right\}^{1/2}
	\end{equation*}
	for a sufficiently large constant $C$ with probability tending to one as $n \to \infty$.
	\label{theorem:singleindex}
\end{theorem}

For the adaptive CHOMP estimator, let $\rho_n = \min_{j=1,\ldots, d} \min_{k \in S_j} \vert \beta_{jk} \vert$, the smallest magnitude of non-zero component across all dimensions. As $n$ grows, we allow $\rho_n$ to converge to a positive finite constant or to zero at a relatively slow rate. Specially, we assume 

\begin{enumerate}[label=(A\arabic*)]	
	\item[(C4)] For each dimension $j=1,\ldots, d$, the initial estimator $\bar{\bbeta}_j$  satisfies $\norm{\bar{\bbeta}_j-\tilde{\bbeta}_j}_\infty = O_p(\delta_n)$ for some sequence $\delta_n \to 0$ such that $\delta_n = o(\rho_n)$.  
	\item[(C5)] (Mutual incoherence) There exists a constant $C$ such that
	$$
	\norm{\mathcal{X}^\top_{S^c} \mathcal{X}_S \left(\mathcal{X}_S^\top \mathcal{X}_S\right)^{-1}}_\infty \leq C < \infty.
	$$ 
	where the notation $\mathcal{X}_S$ denotes the submatrix of $\mathcal{X}$ whose column indices belong to $S$.
	\label{assumption:A4}
\end{enumerate}
Condition (C4) regarding the initial estimator is critical to ensure the weight vector is appropriately defined such that the weights for non-zero coefficients converge to a finite constant, and the weights for the zero coefficients diverge to infinity as the sample size grows. Similar conditions for the initial estimator have been used extensively in the analysis of the Adaptive Lasso for high-dimensional sparse linear models, such as in \cite{zou2006adaptive} and \citet{huang2008adaptive}.  The mutual incoherence condition (C5), which is also commonly used in the analysis of the Adaptive Lasso, is a relatively weak condition on the correlatedness between the active and non-active covariates. With these conditions, we establish the selection consistency of the adaptive CHOMP estimator.
\begin{theorem}
	\label{theorem2}
	Consider a multiple index model with $n\lambda_d = p^\nu$ for $\nu > 1$. Assume conditions \ref{assumption:A1}-(C5) hold, and the number of dimensions $d$ is known. For each dimension $j=1,\ldots,d$ assume 
	$$
	\rho_n^{-\gamma}s^{3/2}n^{-1/2} \rightarrow 0, ~\delta_n^\gamma/\mu_j  \to 0,  ~ n^{-1}\rho_n^{-2\gamma} s\log(p) \rightarrow 0;  ~\rho_n^{-2\gamma} \mu_j s^{1/2} \to 0.
	$$
	Then, the adaptive CHOMP estimator $\hat{ \bbeta}_j^*$ defined in \eqref{eq:adaCholesky matrix penalization} is selection consistent:  $\text{pr}(\hat\beta^*_{jk} \neq 0) \rightarrow 1$ if $k \in S_j$ and $ \text{pr}(\hat\beta^*_{jk} = 0) \rightarrow 1$ if $k \notin S_j$. Furthermore, if $s/n \rightarrow 0$, then the projection matrix $\mathcal{P}(\widehat{\bB^*})$ associated with the adaptive Cholesky matrix estimator satisfies
	$$
	\left\|\mathcal{P}(\widehat{ {\bB}^*})-\mathcal{P}( {\bB}) \right\|_F \leq C \left\{\frac{s \log (p)}{n \lambda_d}\right\}^{1/2}
	$$
	for a sufficiently large constant $C$ with probability tending to one as $n \to \infty$.
\end{theorem}
When the initial estimator $\bar{\bbeta}_j$ is the unpenalized estimate, the quantity $\delta_n = (p/n)^{1/2} \rightarrow 0$. If $\rho_n = O(1)$, Theorem 2 implies selection consistency holds if $s = O(n^\alpha)$ with $\alpha < 1/3$, $s\log(p)/n\rightarrow 0$, and the tuning parameter $\mu_j = O(n^\zeta)$ with $\zeta \in [\gamma(\nu^{-1} - 1), -\alpha/2]$.


Next, we study the large-sample properties of using PIC to select the tuning parameters $\mu_j$ for the adaptive CHOMP estimator. To facilitate theoretical analysis, we study a generalized form of the PIC defined as
\begin{equation}
\textsc{PIC}(\mu_j; \tau_j) = \begin{cases}
	\norm{\mathcal{P}\left\{\hat{ \bbeta}_j({\mu_j})\right\}-\mathcal{P}(\bar{ \bbeta}_j)}_F^2 + \tau_j \norm{\hat{ \bbeta}_j({\mu_j})}_0,& ~ \text{if}~ \hat{ \bbeta}_j({\mu_j}) \neq  \bm{0} \\
	\infty, & ~ \text{if} ~ \hat{ \bbeta}_j({\mu_j}) =  \bm{0},
\end{cases}
\label{eq:PIC_generalized}
\end{equation}
where $\tau_j > 0$ is a model complexity term. Now, for a given value of the tuning parameter $\mu_j$, let $\hat{\bm{\beta}}_j^*({\mu}_j)$ be the corresponding solution of the minimization problem \eqref{eq:adaCholesky matrix penalization} and $\hat{S}_j\left(\mu_j\right) = \{k: \hat{\beta}^*_{jk}(\mu_j) \neq 0\}$. Next, we establish the following result regarding selection consistency of PIC.

\begin{theorem}
		Consider the multiple index model with  the same conditions as in the Theorem \ref{theorem2}. For each dimension $j = 1,\ldots,d$,  denote \[\xi_j = \min \left\{ \dfrac{{\beta}_{jk}^2}{ \bbeta_j^\top \bbeta_j}, \beta_{jk}\neq 0 \right\}
		\] and assume that $\xi_j$ goes to zero at a slower rate than $p/n$. For any sequence $\tau_j$ that goes to zero at a rate slower than $p/n$ but faster than $\xi_j$, i.e  $p/n \stackrel{<}{\sim} \tau_j \stackrel{<}{\sim} \xi_j$,  the adaptive CHOMP estimator with tuning parameter $\mu_j$ selected by minimizing $\textsc{PIC}(\mu_j; \tau_j)$ defined in \eqref{eq:PIC_generalized} with the initial estimate $\bar\bbeta_j$ being the unpenalized estimate satisfies $\text{pr}\left\{\hat{S}_j\left(\mu_j\right)=S_j \right\} \rightarrow 1$ as $n \to \infty$.
		\label{theorem3}
	\end{theorem}

In Theorem \ref{theorem3}, the quantity $\xi_j$ controls the relative magnitude of the minimum non-zero coefficient compared to the $\ell_2$ norm of the $j$th dimension. The condition that the model complexity term $\tau_j$ goes to zero faster than $\xi_j$ ensures minimizing \textsc{PIC} does not lead to underfitting; in other words, when $\xi_j$ is small, the model complexity term $\tau_j$ has to be small as well. Furthermore, 
the term $\tau_j$ has to go to zero at a rate that is slower than $p/n$ to avoid overfitting. If all the non-zero components of $ \bbeta_j$ have the same magnitude, i.e. $\xi_j = o(s_j^{-1})$, then Theorem 3 implies that the rate of convergence to zero is between $p/n$ and $s_j^{-1}$, so we can set $\tau_j = \log(p)/p$ as defined in equation \eqref{first}. In the simulation below, we verify that this choice of $\tau_j$ leads to strong variable selection in finite sample settings. \bred{As far as we are aware, our proposed PIC is the first data-driven approach to select the regularization parameter that theoretically guarantees to achieve variable selection consistency for the central subspace.}

\section{Simulation Studies}
\label{sim}

\subsection{Single index model}
\label{low_dim_sim}
We conduct simulation studies to investigate the performance of the proposed estimators in finite sample settings. In all the settings below, the number of true dimensions $d$ is assumed to be known. For the first simulation, we generate data pairs $( \bm{x}_i^\top, y_i)$ from one of the following models: (I) $  y_i=\sin(\bm{x}_i^\top{\bbeta}_0)\exp(\bm{x}_i^\top{\bbeta}_0)+\varepsilon_i$, (II) $ y_i = 0.5(\bm{x}_i^\top {\bbeta}_0)^3 + \varepsilon_i$, and  (III) $ y_i = \exp(\bm{x}_i^\top\bbeta_0 + \varepsilon_i)$. These models are also considered by \cite{Linsparse} in their simulation study of the single index model. Each row vector $\bm{x}_i$ is independently generated from a $p$-variate Gaussian distribution with mean zero and covariance matrix ${\bSigma} = \mathbf{D} \tilde{\bm\Omega} \mathbf{D}$, where ${\tilde{\bm\Omega}} =  (\tilde\sigma)_{ij}$ is a correlation matrix with elements being either (a) $\tilde\sigma_{ij}=0.5^{|i-j|}$ (autoregressive structure) or type (b) $\tilde \sigma_{ii} = 1$ and $\tilde\sigma_{ij} = 0.5$ when $i \neq j$ (homogeneous structure), and $\mathbf{D}$ is a diagonal matrix whose elements are randomly generated from the uniform distribution Unif$(0.5, 2)$. As a result, each covariate has a different variance.  Next, the vector $ {\bbeta}_0$ is generated with the first $s=5$ components being non-zero. These non-zero components have random sign and magnitude generated from the uniform distribution $\text{Unif}(1,1.5)$. Finally, each random noise term $\varepsilon_i$ is generated independently from the standard normal distribution. The sample size is fixed at $n=1000$ as in \citep{Linsparse},  while the number of covariates varies over  $p \in \{100, 200\}$. For each combination of above parameters, $500$ samples are generated. We set the number of slices to be $H=20$ when computing all the estimators as in \cite{Linsparse}.

We compare the performance of the Matrix Lasso, CHOMP, adaptive CHOMP with $\gamma=1$ and $\gamma=2$, and the Lasso SIR estimators. For the first three estimators, the tuning parameters are selected based on PIC proposed in Section \ref{tuning}. \bred{We use PIC to select the tuning parameter for the Matrix Lasso because  \cite{Linsparse} did not provide a method to select its tuning parameter. How to best select the tuning parameter for the Matrix Lasso is outside the scope of this paper.} The Lasso SIR estimator is implemented with the tuning parameter chosen by ten-fold cross-validation. In the Supplementary Material, we demonstrate that the Lasso SIR with tuning parameter selected via ten-fold cross-validation has roughly the same performance as the Lasso SIR estimator with tuning parameter chosen to minimize the actual estimation error. The latter is not available in practice, because it requires knowledge of the true projection matrix $\mathcal{P}( \bbeta_0)$. We use three metrics for assessing the performance of the estimators, including the estimation error $\norm{\mathcal{P}(\bbeta_0) - \mathcal{P}(\hat{\bbeta})}_F$, the false positive rate (FPR), and the false negative rate (FNR). These metrics are averaged across the $500$ samples. The results for the simulation settings when $\tilde{\bm{\Omega}}$ has the autoregressive structure are presented in Table \ref{tab:tab2}; the results for the settings when $\tilde{\bm{\Omega}}$ has the homogeneous structure show similar conclusions and are presented in section E of the Supplementary Material. 
\begin{table}[!t]
	\centering
	\caption{Performance of the estimators in the single index model simulation with the correlation matrix $\tilde{ \bm{\Omega}}$ having autoregressive structure. Standard errors are included in parentheses. The lowest estimation error for each setting is highlighted. }
	\resizebox{\linewidth}{!}{\begin{tabular}{lllccccc}
			\toprule[1.5pt]
		Model & $p$ & Metric & CHOMP & \multicolumn{2}{c}{Adaptive CHOMP} & Lasso SIR & MLasso \\
		&&&& $\gamma=1$& $\gamma=2$ &&  \\ 
		\hline \addlinespace
		(I) & 100 & Error & 0.26 (0.11) & 0.11 (0.05) & \textbf{0.09} (0.05) & 0.18 (0.06) & 0.57 (0.21) \\ 
		&  & FPR & 0.00 (0.01) & 0.00 (0.00) & 0.00 (0.01) & 0.20 (0.10) & 0.08 (0.08) \\ 
		&  & FNR & 0.00 (0.02) & 0.00 (0.00) & 0.00 (0.00) & 0.00 (0.01) & 0.10 (0.14) \\ 
		& 200 & Error & 0.28 (0.12) & 0.12 (0.08) & \textbf{0.11} (0.05) & 0.20 (0.06) & 0.59 (0.20) \\ 
		&  & FPR & 0.00 (0.01) & 0.01 (0.01) & 0.01 (0.02) & 0.13 (0.07) & 0.06 (0.05) \\ 
		&  & FNR & 0.01 (0.05) & 0.00 (0.00) & 0.00 (0.00) & 0.00 (0.00) & 0.09 (0.12) \\[3pt] 
		
		(II) & 100 & Error & 0.08 (0.11) & \textbf{0.03} (0.01) & \textbf{0.03} (0.01) & 0.05 (0.02) & 0.43 (0.22) \\ 
		&  & FPR & 0.00 (0.00) & 0.00 (0.00) & 0.00 (0.00) & 0.19 (0.10) & 0.03 (0.04) \\ 
		&  & FNR & 0.00 (0.01) & 0.00 (0.00) & 0.00 (0.00) & 0.00 (0.00) & 0.05 (0.09) \\ 
		& 200 & Error & 0.08 (0.15) & \textbf{0.03} (0.01) & \textbf{0.03} (0.01) & 0.06 (0.02) & 0.47 (0.22) \\ 
		&  & FPR & 0.00 (0.00) & 0.00 (0.00) & 0.00 (0.00) & 0.13 (0.06) & 0.02 (0.02) \\ 
		&  & FNR & 0.01 (0.04) & 0.00 (0.00) & 0.00 (0.00) & 0.00 (0.00) & 0.05 (0.09) \\[3pt] 
		
		(III) & 100 & Error & 0.11 (0.10) & \textbf{0.04} (0.02) & \textbf{0.04} (0.02) & 0.07 (0.02) & 0.45 (0.22) \\ 
		&  & FPR & 0.00 (0.00) & 0.00 (0.00) & 0.00 (0.00) & 0.20 (0.10) & 0.04 (0.05) \\ 
		&  & FNR & 0.00 (0.02) & 0.00 (0.00) & 0.00 (0.00) & 0.00 (0.00) & 0.05 (0.10) \\ 
		& 200 & Error & 0.12 (0.16) & \textbf{0.04} (0.02) & \textbf{0.04} (0.02) & 0.08 (0.02) & 0.49 (0.21) \\ 
		&  & FPR & 0.00 (0.00) & 0.00 (0.00) & 0.00 (0.00) & 0.13 (0.06) & 0.02 (0.03) \\ 
		&  & FNR & 0.01 (0.04) & 0.00 (0.00) & 0.00 (0.00) & 0.00 (0.00) & 0.06 (0.10) \\ 
		\bottomrule[1.5pt]
	\end{tabular}}
	\label{tab:tab2}
\end{table}

Tables \ref{tab:tab2} shows that 
the adaptive CHOMP estimator consistently has the best performance in terms of all three metrics. 
In particular, the estimation error of the CHOMP with tuning parameter selected using the PIC is much lower than the estimation error of the Matrix Lasso estimator; these numerical results confirm the benefit of using the Cholesky decomposition. This benefit is strengthened further by the adaptive CHOMP; both the adaptive estimators with $\gamma =1$ and  $\gamma=2$ have the smallest estimation error in all the settings. Regarding variable selection, the Matrix Lasso estimator tends to underfit, while the Lasso SIR estimator tends to overfit. The adaptive CHOMP estimator is able to fully recover the sparsity pattern of $\bbeta_0$ with the average FPRs and FNRs being zero in all the settings. 

\subsection{Multiple index model}
\label{multiple_index_sim}
For the multiple index model, we generate independent data pairs $( {x}_i^\top, y_i)$ from the model (IV) $y_i = (\bm{x}_i^\top \bbeta_1) \left\{\exp( \bm{x}_i^\top \bbeta_2)+\varepsilon_i\right\}, i=1,\ldots, n.$ The model is also considered by \cite{Linsparse} in their simulation study of the multiple index model. The predictors $ \bm{x}_i$ and the random noise $\varepsilon_i$ are generated in the same manner as in Section \ref{low_dim_sim}. We consider two different sparsity patterns in $ {\bbeta}_1$ and $ {\bbeta}_2$. In the first case, the two vectors have the same sparsity patterns; specifically, both of them had the first $s_1=s_2=5$ components nonzero. In the second case, the two vectors have different but overlapping sparsity pattern. Specifically, the first $s_1=5$ components of $ {\bbeta}_1$ are non-zero, while the  fourth to the seventh component of $ {\bbeta}_2$ are non-zero ($s_2 = 4$). The non-zero components are generated in the same way as the non-zero components of $\bbeta_0$ in the single index model simulation. 

For each sample, we compute the same estimators for the first and second dimension separately. All the other parameters, including the tuning parameters for the estimators, are chosen in the same way as in Section \ref{low_dim_sim}. We assess the estimators based on estimation error of the projection matrix, FPR and FNR. Similar to \cite{tan2018convex} then, for the multiple index model, the FPR and FNR  are assessed based on the diagonal of the projection matrix. For example, a false positive for the $j^{th}$ component is counted when $\mathcal{P}(\hat{ {\bB}})_{jj}$ is non-zero but $\mathcal{P}({ {\bB}})_{jj}$ is zero. 
\begin{table}[t]
	\centering
	\caption{Performance of estimators in the multiple index model simulation. Standard errors are in parentheses. The lowest estimation error for each setting is highlighted. }
	\resizebox{\linewidth}{!}{\begin{tabular}{lll ccccc}
				\toprule[1.5pt]
		$p$ & Sparsity & Metric & CHOMP & \multicolumn{2}{c}{Adaptive CHOMP} & Lasso SIR & MLasso \\
		&&&& $\gamma=1$& $\gamma=2$ & & \\ 
		\hline \addlinespace
		100 & Same & Error & 0.33 (0.27) & \textbf{0.22} (0.29) & \textbf{0.22} (0.29) & 0.28 (0.27) & 0.57 (0.26) \\ 
		&  & FPR & 0.00 (0.01) & 0.00 (0.02) & 0.01 (0.02) & 0.32 (0.11) & 0.14 (0.13) \\ 
		&  & FNR & 0.00 (0.00) & 0.00 (0.00) & 0.00 (0.00) & 0.00 (0.00) & 0.03 (0.13) \\ 
		
		& Different & Error & 0.39 (0.13) & 0.24 (0.10) & \textbf{0.22} (0.10) & 0.26 (0.06) & 0.58 (0.26) \\ 
		&  & FPR & 0.00 (0.01) & 0.00 (0.00) & 0.00 (0.00) & 0.39 (0.11) & 0.11 (0.10) \\ 
		&  & FNR & 0.00 (0.02) & 0.00 (0.01) & 0.00 (0.01) & 0.00 (0.00) & 0.02 (0.12) \\[3pt]
		
		200 & Same & Error & 0.33 (0.27) & \textbf{0.22} (0.28) & \textbf{0.22} (0.28) & 0.30 (0.26) & 0.69 (0.19) \\ 
		&  & FPR & 0.00 (0.01) & 0.00 (0.01) & 0.00 (0.01) & 0.22 (0.08) & 0.18 (0.11) \\ 
		&  & FNR & 0.00 (0.02) & 0.00 (0.01) & 0.00 (0.01) & 0.00 (0.01) & 0.04 (0.09) \\ 
		
		& Different & Error & 0.42 (0.13) & 0.24 (0.09) & \textbf{0.21} (0.08) & 0.30 (0.07) & 0.71 (0.17) \\ 
		&  & FPR & 0.00 (0.01) & 0.00 (0.00) & 0.00 (0.00) & 0.27 (0.09) & 0.16 (0.09) \\ 
		&  & FNR & 0.01 (0.02) & 0.00 (0.00) & 0.00 (0.00) & 0.00 (0.00) & 0.03 (0.08) \\ 
				\bottomrule[1.5pt]
	\end{tabular}}
	\label{tab:tab4}
\end{table} 

Table \ref{tab:tab4} demonstrate that the adaptive CHOMP estimator with $\gamma=2$ has the overall best performance among the considered estimators. Similar to the single index model, the CHOMP has considerably smaller estimation error than the Matrix Lasso. Regarding variable selection, as in the single index simulation, the Lasso SIR estimator tends to overfit with too many false positives, whereas the Matrix Lasso estimators tend to underfit with too many false negatives. The CHOMP and adaptive CHOMP estimators are able to recover all the important covariates in two dimensions by having both average false positive rates and false negative rates zero or very close to zero in all the considered settings. 

\bred{In summary, the simulation studies both verify the theoretical results, and demonstrate the superior performance of the adaptive CHOMP estimator, in conjunction with the PIC, in terms of both estimation error and variable selection for the central subspace. This result also confirms a main advantage of the CHOMP approach compared to the Lasso SIR, in that it can be easily generalized to an adaptive version that is able to achieve both estimation and variable selection consistency.}

\section{Data Application}
We apply the methods to a Cancer Trial dataset that contains information about the mean cancer mortality rate and $33$ socioeconomic variables over the 2010-2016 period for $n = 3047$ counties in the United States. The dataset was created from merging several data and is publicly available at \url{https://data.world/nrippner/cancer-trials}. It is of interest to model the mean cancer mortality rate $(y)$ from all these other variables. For illustration purposes, we remove one interval censored predictor (\textit{binnedInc}) that represents the median income per capita binned by decile, and three other predictors having a considerable degree of missingness. This leaves us with $p = 28$ covariates ($\mathbf{X}$), all of which are then standardized before the analysis. 

First, we use the \texttt{dr} package in R to compute the (unpenalized) SIR estimator and estimate the number of dimensions using the chi-square marginal dimension test of \citet{cook2004testing}. The number of slices is set to be $H=20$.  As a result, the number of dimensions of the central subspace is estimated to be 3. Next, we calculate the CHOMP, the adaptive CHOMP with $\gamma = 1$ and the adaptive CHOMP with $\gamma = 2$, and the Lasso SIR estimators. The tuning parameters for these penalized estimators are selected in the same fashion as in the simulation study. Because the true coefficient is not available for real data, we use the distance correlation between sufficient predictors $\mathbf{X}\hat{\mathbf{B}}$ and the response $y$ as a performance measure of the methods, where $\hat{\mathbf{B}}$ is a $ 28 \times 3$ estimated matrix of the three dimensions. A higher distance correlation means a stronger association (both linear and nonlinear) between two variables, thus implying a better prediction ability, see  \citet{szekely2007measuring} and \citet{wang2018principal}. We also examine the number of covariates selected by each method across all the three dimensions.  
\begin{table}[!t]
	\centering
	\caption{Performance of SDR methods in the cancer trial dataset}
	\begin{tabular}{lcc}
		\toprule[1.5pt]
		Methods & Distance correlation & \# of important
		 variables \\ 
		 \hline \addlinespace
		Unpenalized SIR & .59 & 28\\
		CHOMP & .61 & 4 \\
		Adaptive CHOMP ($\gamma = 1$) &.59 & 4  \\
		Adaptive CHOMP ($\gamma = 2$) &.66 & 16 \\
		Lasso SIR &.42 & 28 \\
		\bottomrule[1.5pt]
	\end{tabular}
\label{tab:data_analysis}
\end{table}

\begin{figure}[t]
	\includegraphics[width = \textwidth]{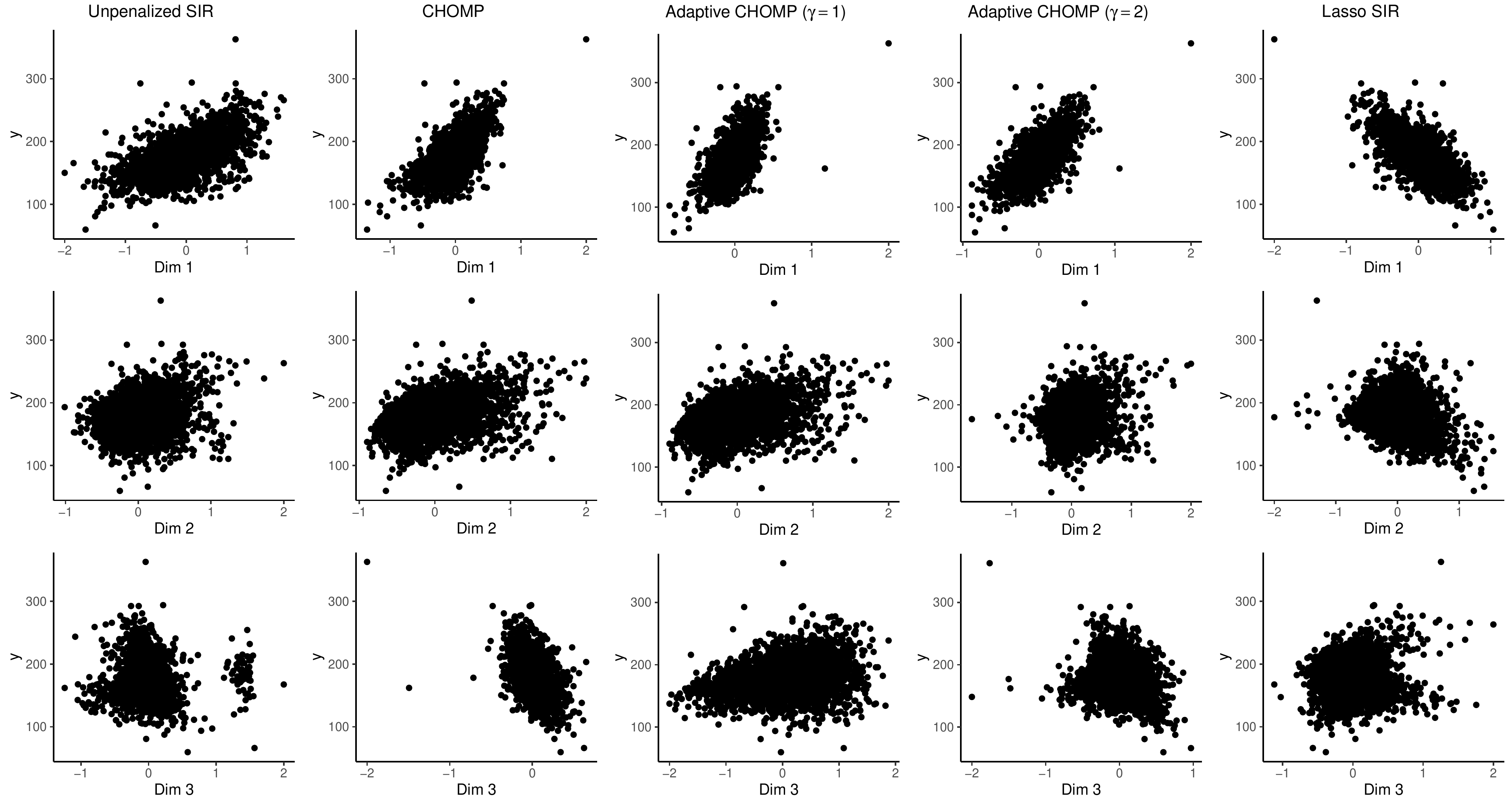}
	\caption{Plots of the response versus each sufficient  predictor obtained by each method in the Cancer Trial dataset application.}
	\label{fig:cancer_data}
\end{figure}

Figure \ref{fig:cancer_data} demonstrates that different methods produce similar sufficient dimensions for the first two dimensions. The response appears to have a strong linear relationship with the first sufficient predictor, while the relationship between the response and the second sufficient predictor is more varied. For the third dimension, different methods produce sufficient estimators with quite different relationships with the response. Table \ref{tab:data_analysis} shows that the adaptive CHOMP estimator with $\gamma =2$ produces sufficient predictors that have highest distance correlation with the response, while the Lasso SIR leads to sufficient predictors with the lowest distance correlation. Regarding variable selection, on the one hand, the Lasso SIR estimator selects all the covariates. As seen in the simulation results, the Lasso SIR estimates tend to have a high FPR; this is likely to happen in this data application as well. On the other hand, the CHOMP estimator and adaptive CHOMP estimator with $\gamma=1$ produce very sparse estimates with only $4$ covariates across all three dimensions; the adaptive CHOMP estimator with $\gamma = 2$ selects $16$ estimates. There are three covariates that are selected by all three estimators: the mean number of diagnoses per capita, poverty rate, and percentage of residents whose highest education level attained is bachelor degree or higher. Compared with simulation results, the performance of the adaptive CHOMP estimator in the data application is more sensitive to the choice of $\gamma$; this may be because the true non-zero coefficients may have wider spread than in the simulation study. The optimal choice for $\gamma$ is a topic of future research.

\section{CHOMP for other inverse regression methods}
\label{sir-otherir}
As discussed at the end of Section \ref{methods}, one advantage of the (adaptive) CHOMP estimator is its ability to extend to other sufficient dimension reduction methods.
For example, we consider a class of methods that satisfy the population equation
$
\bSigma \text{col}(\bB) = \text{col}(\mathbf{Q}),
$
where the matrix $\mathbf{Q}$ is a method-specific kernel matrix. For example,  SIR corresponds to $\mathbf{Q} = \text{var}\{\mathbb{E}( \bX | y) \}$;  sliced average variance estimate (SAVE) corresponds to $\mathbf{Q}  = \mathbb{E}\left\{\bm\Sigma-\text{var}(\bX | y) \right\}^2$;  principal Hessian direction (pHd) corresponds to $\mathbf{Q}  = \mathbb{E}\left[\bX\bX^\top \left\{y - \mathbb{E}(y)\right\}\right]$, among many others. Next, let $\boldeta_1, \ldots, \boldeta_d$ be the eigenvectors associated with the $d$ largest eigenvalues of the kernel matrix $\mathbf{Q}$, then the $j$th method-specific sufficient dimension satisfies $\bSigma \bbeta_j = \boldeta_j, ~ j =1,\ldots, d$. Following the same argument as in Section 3.1, we can then calculate $\hat\bkappa_j$ such that $\hat{\bm{L}}\hat\bkappa_j = \hat{\boldeta}_j$,
where $\hat{\boldeta}_j$ is the $j$th eigenvector of the sample counterpart of the $\textbf{Q}$ matrix respectively. The CHOMP estimator corresponding to each sufficient dimension reduction method can then be constructed as the solution to the minimization  problem \eqref{eq:Cholesky matrix  penalization}, and similarly the adaptive CHOMP estimator is the solution of the minimization problem \eqref{eq:adaCholesky matrix  penalization},  where the weights are set to be $\omega_{jk} = \vert \bar{\beta}_{jk}\vert^{-\gamma}$ with $\bar{\beta}_{jk}$ being the $k$th element of the unpenalized method-specific estimator $\bar{\bbeta}_j = \hat{\bSigma}^{-1}\hat{\boldeta}_j$. We refer to the resulting estimators as, for example, CHOMP-SAVE and adaptive CHOMP-SAVE when the CHOMP and adaptive CHOMP are applied to SAVE; similar definitions hold for pHd.

We conduct a simulation study to demonstrate the performance of these estimators in scenarios where SIR is unable to estimate the sufficient dimension. One such common scenario is when the link function $f$ in \eqref{eq:sufficient dimension reduction} is symmetric around zero \citep[Section 3.2]{li2018sufficient}. We generate data from the single index model (V) $
y_i = (\bm{x}_i^\top \bm\beta_0)^2 + \varepsilon_i
$ and the multiple index model (VI) $y_i = (\bm{x}_i^\top \bm\beta_1)^2 - (\bm{x}_i^\top \bm\beta_2)^4 + \varepsilon_i, ~i=1,\ldots, n$. Each row vector $\bm{x}_i$ is generated from a multivariate normal distribution with the autoregressive correlation structure outlined in Section \ref{low_dim_sim}, and the random noise $\varepsilon_i$ is generated from the standard normal distribution. The vector $\bm{\beta}_0$ is generated as in Section \ref{low_dim_sim}, while the vectors $\bm{\beta}_1$ and $\bm{\beta}_2$ are generated to have different sparsity pattern as in the multiple index model simulation in Section \ref{multiple_index_sim}. We set the sample size to be $n=1000$ and the number of covariates to be $p=100$. We consider the Lasso SIR, CHOMP-SIR, CHOMP-SAVE and CHOMP-pHd estimators, and the adaptive version with $\gamma = 2$ for each CHOMP-based estimator. We use PIC to select the tuning parameters for all the (adaptive) CHOMP estimators. We run each setting on 500 samples and compare the estimators based on the same performance metrics as in Section \ref{sim}.
\begin{table}
	\centering
	\caption{Performance of sufficient dimension estimators in the single index simulation when the true link function $f$ is symmetric. Standard errors are included in parentheses. The lowest estimation error is highlighted for each setting.}
	\resizebox{\textwidth}{!}{\begin{tabular}{rl l lll lll}
		\toprule[1.5pt]
		Model & Metric & Lasso SIR & \multicolumn{3}{c}{CHOMP} &  \multicolumn{3}{c}{Adaptive CHOMP ($\gamma = 2$)}\\
		&&& SIR & SAVE & pHd & SIR & SAVE & pHd \\ 
		\hline \addlinespace
		(V) & Error & 1.41 (0.03) & 1.40 (0.03) & 0.88 (0.35) & 0.54 (0.22) & 1.41 (0.02) & 0.60 (0.45) & \textbf{0.29} (0.20) \\ 
		& FPR & 0.18 (0.11) & 0.03 (0.03) & 0.02 (0.03) & 0.01 (0.02) & 0.09 (0.05) & 0.02 (0.04) & 0.01 (0.02) \\ 
		& FNR & 0.82 (0.20) & 0.96 (0.09) & 0.36 (0.37) & 0.08 (0.15) & 0.90 (0.14) & 0.20 (0.35) & 0.02 (0.10) \\[8pt] 
		(VI) & Error & 1.99 (0.05) & 2.00 (0.05) & 0.84 (0.28) & 1.57 (0.12) & 1.99 (0.02) & \textbf{0.66} (0.32) & 1.54 (0.12) \\ 
		& FPR & 0.27 (0.13) & 0.06 (0.05) & 0.01 (0.03) & 0.03 (0.05) & 0.16 (0.06) & 0.02 (0.04) & 0.11 (0.06) \\ 
		& FNR & 0.71 (0.22) & 1.00 (0.11) & 0.00 (0.13) & 0.57 (0.15) & 0.86 (0.15) & 0.00 (0.09) & 0.43 (0.11) \\ 
		\bottomrule[1.5pt]
\end{tabular}}
\label{tab:otherIR}
\end{table}

Table \ref{tab:otherIR} shows that the SIR-based estimators generally do not perform well in either estimation or variable selection when the true link function $f$ is symmetric around zero. Using CHOMP combined with SAVE or pHd leads to substantially smaller estimation error and improved variable selection. Furthermore, the adaptive CHOMP-SAVE and CHOMP-pHd estimators improve on the performance of the corresponding non-adaptive estimator. For the single index model (V), these adaptive estimators have both low FPRs and FNRs, and the adaptive CHOMP-pHd with $\gamma=2$ has the lowest estimation error. For the multiple index model (VI), the adaptive CHOMP-SAVE estimator with $\gamma=2$ has the lowest estimation error and performs the best in variable selection, while the CHOMP-pHd estimators have a relatively high false negative rate. \bred{All in all, these results demonstrate that the CHOMP can be easily generalized to other sufficient dimension reduction methods, and further confirm the advantages of the adaptive CHOMP-type approach for sparse estimation of the central subspace.}

\section{Discussion}
\bred{This paper presents three main contributions to the literature of sparse sufficient dimension reduction. First, we introduce the CHOMP approach, which is based on the Cholesky decomposition of the sample covariance matrix, for sliced inverse regression estimation of the central subspace, along with the first data-driven projection information criterion theoretically guaranteed to achieve variable selection consistency. Second, though the CHOMP estimator alone may not be as good as the Lasso SIR in simulation studies, the CHOMP approach can be easily generalized to an adaptive version that not only achieves estimation and variable selection consistency, but also has superior performance to the Lasso SIR. Finally, the CHOMP approach is easily extended to other inverse-regression based estimators, for which the corresponding adaptive CHOMP estimators show superior performance in both estimation and variable selection empirically.}

In this paper, we focus on the CHOMP estimators when $n >p$ and $p/n \to 0$ as $n \to \infty$. In this setting, the sample covariance matrix $\hat{\bm{\Sigma}}$ is positive definite and invertible, so is its Cholesky factor $\hat{  \bm{L}}$. In high dimensional settings when $n<p$, the main challenge for using CHOMP is how to estimate $\bm{L}$ given the matrix $\hat{\bm{\Sigma}}$ is no longer positive definite and invertible. In Section D of the Supplementary Material, we explore the use of CHOMP in a high dimensional setting where the Cholesky factor $\bm{L}$ can be estimated efficiently from regression techniques. Future research may investigate the theoretical properties of the CHOMP estimator in such high dimensional settings as well as when combining CHOMP with other sufficient dimension reduction methods. Finally, how to estimate the number of dimensions $d$ from the data in the sparse setting also remains an open problem.


\bibliographystyle{apalike}      
\bibliography{citation1}  

\newpage
\setcounter{section}{0}
\renewcommand\thesection{\Alph{section}}
\renewcommand\thetable{\thesection.\arabic{table}}
\counterwithin{table}{section}

\begin{center}
	\bfseries \Large Supplementary Material
\end{center}
\section{Additional Lemmas and Propositions}
In order to prove the main theoretical results in the main paper, we need additional technical definition and aucillary results.  First, we say that the sample covariance matrix of predictors  $\hat{ {\bSigma}}$ satisfies the restricted eigenvalue condition over a set $T$ with parameter $(q, r)$ if and only if
\begin{equation*}
	\left\|\hat{ \bm{L}}^\top {v}\right\|_2^2 =  {v}^\top \hat{ {\bSigma}} \bm{v} \geq r\left\| \bm{v}\right\|_2^2, \text{ for all }  \bm{v} \in  \mathcal{C}(T, q) = \{ \bm{v} \in \mathbb{R}^p ~|~ \left\| \bm{v}_{T^c}\right\|_1 \leq q \left\|\bm{v}_{T}\right\|_1\}.
\end{equation*} 
This condition is essential in obtaining the consistency of the lasso estimate in the linear model \citep{wainwright2019high}. It is also essential for the consistency of the Cholesky matrix penalization estimate as shown below.  

We begin by obtaining the following bound of the difference between the CHOMP estimate and the pseudo-true parameter as defined in Section 4 of the main paper.
\begin{lemma}
	\label{lemma1}
	Assume the sample covariance matrix $\hat{ {\bSigma}}$
	satisfies the restricted eigenvalue condition with parameter $q=3$ and some positive constant $r$. Then, any solution of the equation (3.8) of the main paper  with tuning parameter bounded below as $\mu_j \geq 2 \norm{\hat{ \boldeta}_j-\hat{ {\bSigma}}\tilde{ \bbeta_j}}_\infty$ satisfies 
	$\left\|\hat{ \bbeta}_j-\tilde{ \bbeta}_j\right\|_2 \leq 3{r}^{-1} \mu_j s_j^{1/2}$, for $j=1,\ldots, d$. 
	\label{eq:deterministic}
\end{lemma}
This result parallels the basic consistency result for the lasso in the linear model (\cite{wainwright2019high}). The bound on the right hand side of \eqref{eq:deterministic} is inversely proportional to the restricted eigenvalue constant $\theta$, which is expected because a higher $\theta$ implies a higher curvature around the optimal $\hat{ {\beta}_j}$. Also, the bound scales with $s_j^{1/2}$; this is also natural because we are trying to estimate an unknown vector with $s_j$ non-zero entries. We first prove the Lemma 1.

\subsection{Proof of Lemma 1}
As the Lemma 1 holds for each dimension $j = 1,\ldots, d$, we remove the subscript $j$ in the development below.  First, we prove that $ \delta \in C(S,3)$ defined in the paper. By definition of $\hat{ {\bbeta}}$, we have
\begin{equation*}
	\dfrac{1}{2}\norm{\hat{ {L}}^\top\hat{ {\bbeta}} - \hat{ {\bkappa}}}_2^2 + \mu \norm{\hat{ {\bbeta}}}_1 \leq \dfrac{1}{2} \norm{\hat{ {L}}^\top\tilde{ {\bbeta}} - \hat{ {\bkappa}}}_2^2 + \mu \norm{\tilde{ {\bbeta}}}_1.
	\label{eq:chompopt1}
\end{equation*}
Writing $\hat{ {\bbeta}} = \tilde{ {\bbeta}}+ {\bdelta}$  we obtain
\begin{equation}
	\dfrac{1}{2}\norm{\hat{ \bm{L}}^\top{ {\bdelta}}-\left( \hat{ {\bkappa}}-\hat{ \bm{L}}^\top\tilde{ {\bbeta}}\right)}_2^2 + \mu \norm{\hat{ {\bbeta}}}_1 \leq \dfrac{1}{2} \norm{\hat{ \bm{L}}^\top\tilde{ {\bbeta}} - \hat{ {\bkappa}}}_2^2 + \mu \norm{\tilde{ {\bbeta}}}_1.
	\label{eq:chompopt2}
\end{equation}
Expanding the first term on the left hand side of \eqref{eq:chompopt2}, we have \[\dfrac{1}{2}\norm{\hat{ \bm{L}}^\top{ {\bdelta}}-\left( \hat{ {\bkappa}}-\hat{ \bm{L}}^\top\tilde{ {\bbeta}}\right)}_2^2 =\dfrac{1}{2} \norm{\hat{ \bm{L}}^\top\tilde{ {\bbeta}} - \hat{ {\bkappa}}}_2^2 -  {\bdelta}^\top\hat{ \bm{L}}(\hat{ {\bkappa}}-\hat{ \bm{L}}^\top\tilde{ {\bbeta}}) + \dfrac{1}{2} \norm{\hat{ \bm{L}}^\top {\bdelta}}_2^2,
\] 
Hence,
\begin{equation}
	\begin{split}
		0 \leq \dfrac{1}{2}\norm{ \bm{L}^\top {\bdelta}}_2^2 & \leq  {\bdelta}^\top\hat{ \bm{L}}(\hat{ {\bkappa}}-\hat{ \bm{L}}^\top\tilde{ {\bbeta}})+\mu\left(\norm{\tilde{ {\bbeta}}}_1-\norm{\hat{ {\bbeta}}}_1\right) \\
		& \stackrel{(i)}\leq \norm{ {\bdelta}}_1\norm{\hat{ \bm{L}}\hat{ {\bkappa}}-\hat{ \bm{L}}\hat{ \bm{L}}^\top\tilde{ {\bbeta}}}_\infty + \mu\left(\norm{\tilde{ {\bbeta}}}_1-\norm{\hat{ {\bbeta}}}_1\right) \\
		& \stackrel{(ii)}{=} \norm{ {\bdelta}}_1 \norm{\hat{ \boldeta}-\hat{ {\bSigma}}\tilde{ \bbeta}}_\infty  + \mu\left(\norm{\tilde{ {\bbeta}}}_1-\norm{\hat{ {\bbeta}}}_1\right) \\ 
		& \stackrel{(iii)}\leq \dfrac{1}{2}\mu\norm{ {\bdelta}}_1  + \mu\left(\norm{\tilde{ {\bbeta}}}_1-\norm{\hat{ {\bbeta}}}_1\right) \label{eq:10},
	\end{split}
\end{equation}
where step $(i)$ follows from Holder's inequality, step $(ii)$ follows from the definitions $\hat{ {\bkappa}} = \hat{ \bm{L}}^{-1}\hat{ {\boldeta}}$ and $\hat{ {\bSigma}}=\hat{ \bm{L}}\hat{ \bm{L}}^\top$, and step $(iii)$ follows from the condition $\mu \geq 2 \norm{\hat{ \boldeta}-\hat{ {\bSigma}}\tilde{ \bbeta}}_\infty.$ Then, we have \begin{equation}\dfrac{1}{2}\norm{ {\bdelta}}_1+\norm{\tilde{ {\bbeta}}}_1-\norm{\hat{ {\bbeta}}}_1 \geq 0. 
	\label{eq:delta}
\end{equation}
Because $\hat{ {\bbeta}} = \tilde{ {\bbeta}}+ {\bdelta}$  and $\tilde{ {\bbeta}}_{T^c} =  {0}$, applying the (reverse) triangle inequality gives 
\begin{align*}
	\norm{\tilde{ {\bbeta}}}_1-\norm{\hat{ {\bbeta}}}_1  = \norm{\tilde{ {\bbeta}}_S}_1 - \norm{\tilde{ {\bbeta}}_S+ {\bdelta}_S}_1 - \norm{ \bdelta_{S^c}}_1 \leq \norm{ {\bdelta}_S}_1 -\norm{ {\bdelta}_{S^c}}_1, 
\end{align*} 
Furthermore, we have $\norm{ {\bdelta}}_1 = \norm{ {\bdelta}_S}_1+\norm{ {\bdelta}_{S^c}}_1$. Therefore, equation \eqref{eq:delta} gives
\begin{align*}
	0 & \leq \dfrac{1}{2}\norm{ {\bdelta}}_1+\norm{\tilde{ {\bbeta}}}_1-\norm{\hat{ {\bbeta}}}_1 \\ & \stackrel{(iv)}{\leq} \dfrac{1}{2}{\norm{ {\bdelta}_S}_1+\dfrac{1}{2}\norm{ {\bdelta}_{S^c}}_1} +\norm{ {\bdelta}_S}_1 -\norm{ {\bdelta}_{S^c}}_1  = \dfrac{3}{2}\norm{ {\bdelta}_{S}}_1 - \dfrac{1}{2} \norm{ {\bdelta}_{S^c}}_1  \\& \stackrel{(v)}{\leq} \dfrac{3}{2}\norm{ {\bdelta}_{S}}_1. 
\end{align*}
It follows from step (iv) that $\norm{ {\bdelta}_{S^c}}_1 \leq 3 \norm{ {\bdelta}_{S}}_1$, or $ \bdelta \in C(S,3)$.	Finally, applying the restricted eigenvalue condition of the sample covariance matrix (defined in section 4 of the main paper), we obtain
\[
\begin{split}
	\dfrac{1}{2}\theta \norm{ \bdelta}_2^2 {\leq} \dfrac{1}{2}\norm{\hat{ \bm{L}}^\top {\bdelta}}_2^2 & \stackrel{(vi)}{\leq} \dfrac{1}{2}\mu\norm{ {\bdelta}}_1  + \mu(\norm{\tilde{ {\bbeta}}}_1-\norm{\hat{ {\bbeta}}}_1)\\& \stackrel{(vii)}{\leq} \dfrac{3}{2} \mu\norm{ \bdelta_S}_1 \stackrel{(viii)}{\leq} \dfrac{3}{2}\mu\sqrt{s}\norm{ \bdelta_S}_2 \leq \dfrac{3}{2}\mu\sqrt{s}\norm{ \bdelta}_2,
\end{split}
\]
where step $(vi)$ follows from \eqref{eq:10}, step $(vii)$ follows from step $(v)$, and step $(viii)$ follows from the Cauchy-Schwartz inequality. Finally, we obtain
\[\norm{\hat{ \bbeta}-\tilde{ \bbeta}}_2 = \norm{ \bdelta}_2 \leq \dfrac{3}{\theta}\mu \sqrt{s}\] as required.

\subsection{Additional Propositions}

Next, we state the following results from \cite{Linsparse} which essentially imply that the conditions for the Lemma 1 in the main paper hold with probability tending to one. We begin with the restricted eigenvalue condition for the sample covariance matrix $\hat{{\bSigma}}$. 
\begin{proposition}
	Assume Condition (C1) in the paper holds. For some universal constants $a_1, a_2$ and $a_3$, if the sample size $n$ satisfies  $n>a_1 s\log(p)$, then the sample covariance matrix $\hat{ \bSigma}$ satisfies the restricted eigenvalue condition with parameter $(q, r) =(3, \sqrt{C_\text{min}}/8)$ over any set $T$ of cardinality $s$ with probability at least $1-a_2 \exp(-a_3n)$.
	\label{prop:prop1}
\end{proposition}
Next, one key condition in the Lemma 1 is that the tuning parameter  has to satisfy the lower bound $\mu \geq 2 \norm{\hat{ \boldeta}-\hat{ {\bSigma}}\tilde{ \bbeta}}_\infty.$ Proposition \ref{prop:prop2} implies that this lower bound is well-controlled.

\begin{proposition}
	\label{prop:prop2}
	Assume conditions (C1)-(C3) in the main paper hold. Then 
	$$
	\norm{\hat{ \boldeta}_j-\hat{{\bSigma}}\tilde{\bbeta}_j}_\infty = O_p\left\{\dfrac{\log(p)^{1/2}}{(n\hat\lambda_j)^{1/2}}\right\}, ~\quad j =1,\ldots, d.
	$$
\end{proposition}
Proposition \ref{prop:prop2} implies that if we set  $\mu_j = M\left\{{\log(p)}/(n\hat{\lambda}_j)\right\}^{1/2}$  for a sufficiently large constant $M$, then we have $\mu_j \geq 2\norm{\hat{ \boldeta}-\hat{ {\bSigma}}\tilde{ \bbeta}}_\infty$ with probability tending to one. When $n\rightarrow \infty$, the ratio $p/n \rightarrow 0$, and hence  $\log(p) /n \rightarrow 0$. As long as the eigenvalue $\hat{\lambda}_j$ is bounded away from zero, $\norm{\hat{ \boldeta}_j-\hat{ {\bSigma}}\tilde{ \bbeta}_j}_\infty \rightarrow 0$, then any positive tuning parameter $\mu_j$ will asymptotically satisfy the bound.
\begin{proposition}
	If $n\lambda = p^\nu$ for $\nu> 1/2$, then 
	$
	\|\tilde{ {\bbeta}}_j\|_2 \geq C \left(\lambda_j/{\hat{\lambda}_j}\right)^{1/2}$ and $  \left(\lambda_j/{\hat{\lambda}_j}\right)^{1/2} \leq C \left\|
	\mathcal{P}( \Lambda) \widehat{ \boldeta}_j\right\|_{2} $ for $ j =1,\ldots, d$	with probability tending to one. 
	\label{prop:prop3}
\end{proposition}

Proposition \ref{prop:prop3} implies that the norm of the pseudo-true parameter $\norm{\tilde{\bbeta}_j}_2$ is bounded away from zero and that the ratio ${\lambda}_j/\hat{\lambda}_j$ is bounded for each $j = 1,\ldots, d$
\section{Proof of Main Theorems}
\subsection{Proof of Theorem 1}
When the sample size $n \rightarrow \infty$, Proposition 1 in the main paper implies that the sample covariance matrix  $\hat{ \bSigma}$ satisfies the restricted eigenvalue  condition with probability tending to one.  Furthermore, the condition on the tuning parameter $\mu$ implies that we can apply Lemma 1 with $\theta = C_\text{min}^{1/2}/8 > 0$. Hence for each dimension $j=1,\ldots,d$, we then have
-\[ \norm{\bdelta_j}_2 =  \norm{\hat{ {\bbeta}}_j-\tilde{ {\bbeta}}_j}_2 \leq {24}C_\text{min}^{-1/2} s_j^{1/2}{\mu}_j \leq C \left\{\dfrac{s_j\log(p)}{n{\lambda_j}}\right\}^{1/2}\leq  C \left\{\dfrac{s\log(p)}{n{\lambda_d}}\right\}^{1/2}. 
\]
Proposition $3$ and Condition (C2) in the main paper imply that the norm  $\norm{\tilde{ {\bbeta}}_j}_2$ is bounded away from zero. As a result, 
\begin{equation*}
	\left\|\mathcal{P}(\widehat{ \bbeta}_j)-\mathcal{P}( \bbeta_{j})\right\|_{F}= \left\|\mathcal{P}(\widehat{ \bbeta}_j)-\mathcal{P}(\tilde{ \bbeta}_j)\right\|_{F}\leq 4 \frac{\|\widehat{{\bbeta}}_j-\widetilde{{\bbeta}}_j\|_{2}}{\|\widetilde{{\bbeta}}\|_{2}}=4\dfrac{\|{\bdelta}_j\|_{2}} {\|\widetilde{{\bbeta}}_j\|_{2}} \leq C\left\{\dfrac{s\log(p)}{n\lambda_d}\right\}^{1/2}
	\label{equation2}
\end{equation*}
for a sufficiently large constant $C$. Furthermore, \cite{Linsparse} shows that the lengths of each vector $\tilde{\bbeta}_j, ~j=1,\ldots, d$ are bounded below by $C(\lambda/\hat{\lambda}_j)^{1/2}$, and the angles between any two vectors of  $\tilde{\bbeta}_j, ~j=1,\ldots, d$ are bounded below by a constant. The Gram-Schmidt process then implies 
$$
\left\|\mathcal{P}(\widehat{ {\bB}})-\mathcal{P}( {\bB}) \right\|_F\leq C \left\{\frac{s \log (p)}{n \lambda}\right\}^{1/2}
$$ as claimed.

\subsection{Proof of Theorem 2}
It suffices to prove selection consistency for each dimension. In the proof below, the notations $\tilde\bbeta$, $\bar\bbeta$, and $\bbeta^*$ denote the pseudo-true parameter (defined in the main paper), the initial consistent estimate, and the Adaptive Cholesky estimate for each dimension respectively; furthermore the set $S$ is the true index set of non-zero components of $\bbeta$. The subscript used in the proof, for example $\bbeta_k$, denotes the $k$th component of $\bbeta$, and $\bbeta_S$, denotes the vector of components of $\bbeta$ whose indices belong to $S$. For any matrix $A$ and a set $T$, the notation $A_{,T}$ and $A_{T,}$ denotes the submatrix of $A$ with column indices in $T$ and the submatrix of $A$ with row indices in $T$ respectively, and $A_{T,T}$ denotes the submatrix with both row and column indices in $T$.

First, let $\hat{\bDelta} = \text{diag}(\bar\bbeta_1^\gamma, \ldots, \bar{\bbeta}_p^\gamma)$, a diagonal matrix whose elements correspond to the inverse of the weight vector $\omega$. For ease of notation, consider the case of $\gamma =1$. Due to consistency of the initial estimator, the matrix $\hat{\bDelta}_{S,S}$ is invertible with probability one. Furthermore, since the sample covariance matrix $\hat{\bSigma}$ satisfies the restricted eigenvalue condition with probability tending to one (Proposition 1), the minimum eigenvalue of the matrix $\hat{\bSigma}_{S,S}$ is bounded away from zero with probability tending to one as well. In that case, each component of the adaptive Cholesky matrix penalization estimator can be computed as $\hat\bbeta^*_k = \bar\bbeta_k\hat{u}_k,$  $k=1,\ldots, p$, where the vector $\hat{\bm{u}}=(\hat{u}_1, \ldots, \hat{u}_p)^\top$ solves the following minimization problem
$$
\hat{\bm{u}} = \arg\min_{\bm{u}} \dfrac{1}{2} \norm{\bm{V} \bm{u} - \hat{ {\bkappa}}}_2^2 + \mu\norm{\bm{u}}_1
$$ 
with $\bm{V} = \hat{\bm{L}}^\top \hat{\bDelta}$. Therefore, if $\hat{\bm{u}}$ recovers the exact sparsity pattern, so does the adaptive Cholesky matrix penalization estimate. From the Karush-Kuhn-Tucker condition, the estimate $\hat{u}$ satisfies 
\begin{equation}
	\bm{V}^\top \bm{V} \hat{\bm{u}} - \bm{V}^\top \hat\bkappa + \mu w = 0,
	\label{kkt}
\end{equation}
where $\bm{w} = (w_1, \ldots, w_p)$ with $w_k = \text{sign}(\hat{u}_k)$ if $\hat{u}_k \neq 0$ and $
\vert w_k \vert \leq 1$ otherwise. Therefore, $\hat{\bm{u}}$ recovers the exact sparsity pattern of $\bbeta^*$  if and only if $\bm{u}_S \neq 0,\bm{w}_S = \text{sign}(\bbeta_S), u_{S^c} = 0, \vert w_{S^c} \vert \leq 1$. Furthermore, by definition of $\bkappa$, the quantity $\bm{V}^\top\hat\bkappa = \hat{\bDelta}^\top \hat{\bm{L}}\hat\bkappa = \hat{\bDelta} \hat{\boldeta}$, and $\bm{V\hat{u}} = \bm{V}_{,S}\hat{\bm{u}}_S$.  Combining these with condition \eqref{kkt} above, if $\hat{\bm{u}}$ recovers the exact sparsity pattern of $\bbeta$, we then have   
$$
\begin{aligned}
	\bm{V}^\top_{S,} \bm{V}_{,S}\hat{\bm{u}}_S - \hat{\bDelta}_{S,S}\hat{\boldeta}_S + \mu\text{sign}(\bbeta_S) & = 0 \\
	\bm{V}^\top_{S^c,} \bm{V}_{,S}\hat{\bm{u}}_S - \hat{\bDelta}_{S^c, S^c}^\top\hat{\boldeta}_{S^c} + \mu \bm{w}_{S^c} & = 0. 
\end{aligned}
$$
Solving this system of equations, we then have
$$
\begin{aligned}
	\hat{\bm{u}}_S &  = (\bm{V}_{S,}^\top \bm{V}_{,S})^{-1} \left\{ \hat{\bDelta}_{S,S}^\top\hat{\boldeta}_S - \mu\text{sign}(\bbeta_S)\right\}   \\
	-\mu \bm{w}_{S^c} & =  \bm{V}^\top_{S^c,} \bm{V}_{,S} (\bm{V}_{S,}^\top \bm{V}_{,S})^{-1} \left\{ \hat{\bDelta}_{S,S}^\top\hat{\boldeta}_S - \mu\text{sign}(\bbeta_S)\right\}   - \hat{\bDelta}_{S^c, S^c}\hat{\boldeta}_{S^c}.
\end{aligned}
$$
(1) With this in mind,  we will show that probability of underselection goes to zero by showing that $\text{pr}(\hat{u}_{S} \neq 0) \to 1$. In fact,
$$
\begin{aligned}
	(\bm{V}_{S,}^\top \bm{V}_{,S})^{-1}
	\hat{\bDelta}_{S,S}^\top\hat{\boldeta}_S  & = \left( \hat{\bDelta}_{S,S} \hat{\bm{L}}_{S,} \hat{\bm{L}}_{,S}^\top \hat{\bDelta}_{S,S}\right)^{-1}\hat{\bDelta}_{S,S} \hat{\boldeta}_S  
	=  \hat{\bDelta}_{S,S} ^{-1} \hat{\bSigma}_{S,S}^{-1}  \hat{\boldeta}_S \\
	& = \hat{\bDelta}_{S,S}^{-1} \hat{\bSigma}_{S,S}^{-1} \left(\hat{\boldeta}_S - \tilde{\boldeta}_S\right) + \hat{\bDelta}_{S,S}^{-1} \hat{\bSigma}_{S,S}^{-1} \tilde{\boldeta}_S \\
	& =  \underbrace{\hat{\bDelta}_{S,S}^{-1} \hat{\bSigma}_{S,S}^{-1} \left(\hat{\boldeta}_S - \tilde{\boldeta}_S\right)}_{I_1} + \underbrace{\hat{\bDelta}_{S,S}^{-1} \hat{\bSigma}_{S,S}^{-1} \bSigma_{S,S}\tilde{\bbeta}_S}_{I_2},
\end{aligned}
$$
where the last inequality follows from the definition that $\tilde\boldeta = \bSigma \tilde\bbeta$ and the vector $\tilde{\bbeta}$ is a sparse vector. Using Proposition 2, condition (C1) and (C4), we then have
$$
\norm{I_1}_\infty \leq \norm{\bDelta_{S,S}^{-1}}_\infty \norm{\hat{\bSigma}_{S,S}^{-1}}_\infty \norm{\hat{\boldeta}_S - \tilde{\boldeta}_S}_\infty \leq  \dfrac{s^{1/2}}{\rho_nC_\text{min}^{1/2}} O\left(\frac{\log(p)^{1/2}}{(n\lambda)^{1/2}} \right) =   O_p \left\{ \frac{s^{1/2}\log(p)^{1/2}}{\rho_n(n\lambda)^{1/2}} \right\} \rightarrow 0,
$$
since $n^{-1}\lambda^{-1}\rho_n^{-2} s\log(p) \rightarrow 0$. Next,  
$$
\begin{aligned}
	I_2 = \hat{\bDelta}_{S,S}^{-1} \hat{\bSigma}_{S,S}^{-1} \bSigma_{S,S}\tilde{\bbeta}_S & = \hat{\bDelta}_{S,S}^{-1}\hat{\bSigma}_{S,S}^{-1} \hat{\bSigma}_{S,S}\tilde{\bbeta}_S + \hat{\bDelta}_{S,S}^{-1}\hat{\bSigma}_{S,S}^{-1}\left(\bSigma_{S,S} - \hat{\bSigma}_{S,S}\right)\hat\bbeta_S \\ &
	= \hat{\bDelta}_{S,S}^{-1} \tilde{\bbeta}_S + \hat{\bDelta}_{S,S}^{-1}\hat{\bSigma}_{S,S}^{-1}\left(\bSigma_{S,S} - \hat{\bSigma}_{S,S}\right)\hat\bbeta_S = I_{21} + I_{22}.
\end{aligned}
$$
Due to the consistency of the initial estimator $\bar{\bbeta}$, each element of the term $I_{21}$ converges to a non-zero constant with probability 1 at a rate $O_p(\bdelta_n)$ since $\bdelta_n = o(\rho_n)$. Since $\norm{\bSigma_{S,S} - \hat{\bSigma}_{S,S}}_2 = O\left\{(s/n)^{1/2}\right\}$ \citep{wainwright2019high},  we have $\norm{\bSigma_{S,S} - \hat{\bSigma}_{S,S}}_\infty = O(sn^{-1/2})$ and
$$
\norm{I_{22}}_\infty \leq \norm{\hat\bDelta_{S,S}^{-1}}_\infty \norm{\hat{\bSigma}_{S,S}}^{-1}_\infty \norm{\bSigma_{S,S} - \hat{\bSigma}_{S,S}}_\infty \norm{\tilde\bbeta_S}_\infty =  O_p\left\{\rho_n^{-1}s^{3/2}n^{-1/2}\right\} \rightarrow 0 $$
since $\rho_n^{-1}s^{3/2}n^{-1/2} \rightarrow 0$. Finally, we consider  $\mu (V_S^\top V_S)^{-1}  \text{sign}(\bbeta_S) = \mu \left(\hat\bDelta_{S,S} \hat{\bSigma}_{S,S} \hat{\bDelta}_{S,S}\right)^{-1} \text{sign}(\bbeta_S)$. We have
$$
\norm{\mu \left(\hat\bDelta_{S,S} \hat{\bSigma}_{S,S} \hat{\bDelta}_{S,S}\right)^{-1} \text{sign}(\bbeta_S)}_\infty \leq \mu O_p(s^{1/2})\norm{\hat\bDelta_{S,S}^{-1}}_\infty^2 = O_p\left(\frac{\mu s^{1/2}}{\rho_n^2}\right)
$$
so this term also goes to zero when $\mu = o(\rho_n^2/s^{1/2})$.

(2) We will show that the probability of overselection also goes to zero. Define the term
$$
Q = \bm{V}^\top_{S^c,} \bm{V}_{,S} (\bm{V}_{S,}^\top \bm{V}_{,S})^{-1} \left\{ \hat{\bDelta}_{S,S}^\top\hat{\boldeta}_S - \mu\text{sign}(\bbeta_S)\right\}   - \hat{\bDelta}_{S^c, S^c}\hat{\boldeta}_{S^c},
$$
so there would be no over-selection if $\norm{Q}_\infty \leq \mu$. By the triangle inequality and the fact that $\norm{\hat{\boldeta}}_{\infty} = \vert \text{sign}({\bbeta_S}) \vert \leq  1$, we have
$$
\begin{aligned}
	\norm{Q}_\infty & \leq \norm{\hat{\bDelta}_{S^c, S^c}\hat{\boldeta}_{S^c}}_\infty + \norm{\bm{V}^\top_{S^c,} \bm{V}_{,S} (\bm{V}_{S,}^\top \bm{V}_{,S})^{-1} \hat{\bDelta}_{S,S}^\top\hat{\boldeta}_S}_\infty + \mu \norm{\bm{V}^\top_{S^c,} \bm{V}_{,S} (\bm{V}_{S,}^\top \bm{V}_{,S})^{-1}\text{sign}(\bbeta_S)}_\infty \\  
	& \leq \norm{\hat{\bDelta}_{S^c, S^c}}_\infty +  \norm{\hat{\bDelta}_{S^c, S^c}\hat{\bm{L}}_{S^c,} \hat{\bm{L}}_{,S}^\top \left( \hat{\bm{L}}_{S,} \hat{\bm{L}}_{,S}^\top  \right)^{-1} }_\infty + \mu \norm{\hat\bDelta_{S^c, S^c} \hat{\bm{L}}_{S^c,} \hat{\bm{L}}_{,S}^\top \left( \hat{L}_{S,} \hat{\bm{L}}_{,S}^\top\right)^{-1} \hat\bDelta_{S,S}^{-1}}_\infty \\ 
	& = O_p(\bdelta_n) + O_p(\bdelta_n) + O_p \left(\frac{\mu\bdelta_n}{\rho_n} \right) \leq \mu 
\end{aligned}
$$
as long as $\bdelta_n/\mu \rightarrow 0$, where the last equality follows from  $\norm{\hat{\bm{L}}_{S^c,} \hat{\bm{L}}_{,S}^\top \left( \hat{\bm{L}}_{S,} \hat{\bm{L}}_{,S}^\top  \right)^{-1}}_\infty = \norm{\mathcal{X}^\top_{S^c}  \mathcal{X}_S \left(\mathcal{X}_S^\top \mathcal{X}_S\right)^{-1}}_\infty = O(1)$ by condition (C5).

(3) Finally, we show the bound on the error of the projection matrix associated with the Adaptive Cholesky Matrix $\hat{B}^*$. By the same argument as in the proof of Theorem 1 (section 1.3), it suffices to show that $\norm{\hat\bbeta^* - \tilde{\bbeta} }_2 \leq C s^{1/2}\log(p)^{1/2}/(n\lambda)$. In fact, due to variable selection consistency, it suffices to show the bound holds for  $\norm{\hat\bbeta_S^* - \tilde{\bbeta}_S }_2$. The first-order condition then implies
$$
\bm{V}^\top_{S,} \bm{V}_{,S} \hat{\bm{u}}_S - \hat{\bDelta}_{S,S}\hat{\boldeta}_S + \mu \text{sign}(\hat{\bm{u}}_S) = 0
$$
By definition, we have $\hat{\bm{u}}_S = \hat{\bDelta}_{S,S}^{-1}\hat{\bbeta}^*_S$ and $\bm{V}_{,S} = \hat{\bm{L}}^\top_{,S} \hat{\bDelta}_{S,S}$, so substituting them into the above equation gives

$$
\begin{aligned}
	& \hat{\bDelta}_{S,S} \hat{\bm{L}}_{S,} \hat{\bm{L}}^\top_{,S}\hat{\bbeta}_S^*  = \hat{\bDelta}_{S,S} \hat{\boldeta}_S - \mu\text{sign}(\hat{\bm{u}}_S) \\
	\text{or ~}  & \hat{\bDelta}_{S,S} \hat{\bSigma}_{S,S}\hat{\bbeta}_S^*  =  \hat{\bDelta}_{S,S}(\hat{\boldeta}_S - \tilde{\boldeta}_S) + \hat{\bDelta}_{S,S}\tilde{\boldeta}_S - \mu\text{sign}(\hat{\bm{u}}_S)
\end{aligned}
$$
Also, by definition $\tilde{\boldeta}_S = \bSigma_{S,S}^{-1}\tilde{\bbeta}_S$, so substituting it into the above equation and doing one algebraic manipulation gives
\begin{align*}
	& \hat{\bDelta}_{S,S} \hat{\bSigma}_{S,S} (\hat\bbeta^*_S - \tilde{\bbeta}_S)   = \hat{\bDelta}_{S,S} (\hat{\boldeta}_S - \tilde{\boldeta}_S)+\hat{\bDelta}_{S,S} (\bSigma_{S,S} - \hat{\bSigma}_{S,S})\tilde{\bbeta}_S - \mu\text{sign}(\hat{\bm{u}}_S) \\ 
	\text{or}\quad &  (\hat\bbeta^*_S - \tilde{\bbeta}_S) = \hat{\bSigma}_{S,S}^{-1} (\hat{\boldeta}_S - \tilde{\boldeta}_S) + \hat{\bSigma}_{S,S}^{-1}(\bSigma_{S,S} - \hat{\bSigma}_{S,S})\tilde{\bbeta}_S  - \mu \hat{\bSigma}_{S,S}^{-1}\hat{\bDelta}_{S,S}^{-1} \text{sign}(\hat{\bm{u}}_S).
\end{align*}
Therefore, the triangular inequality and the fact that $\text{sign}(\hat{u}_S) = \pm1$ gives
$$
\begin{aligned}
	\norm{\hat\bbeta^*_S - \tilde{\bbeta}_S}_\infty & \leq \norm{\hat{\bSigma}_{S,S}^{-1}}_2\norm{\hat{\boldeta}_S - \tilde{\boldeta}_S}_\infty + \norm{\hat{\bSigma}_{S,S}^{-1}}_2\norm{\bSigma_{S,S} - \hat{\bSigma}_{S,S}}_2 \norm{\tilde{\bbeta}_S}_{\infty} + \mu \norm{\hat{\bSigma}_{S,S}^{-1}}_\infty \norm{\hat\bDelta_{S,S}^{-1}}_\infty \\
	& =  O_p\left\{\dfrac{\log(p)^{1/2}}{(n\lambda)^{1/2}}\right\} + O_p\left(\dfrac{s^{1/2}}{n^{1/2}}\right) + O_p(\mu s^{1/2}\rho_n^{-1}) = O_p\left\{\dfrac{\log(p)^{1/2}}{(n\lambda)^{1/2}}\right\}
\end{aligned}
$$
due to the condition of the tuning parameters as stated in the Theorem. Finally, we have 
$$
\norm{\hat\bbeta^*_S - \tilde{\bbeta}_S}_2 \leq s^{1/2}\norm{\hat\bbeta^*_S - \tilde{\bbeta}_S}_\infty \leq C \left\{\dfrac{s\log(p)}{(n\lambda)}\right\}^{1/2}
$$ for a sufficiently large constant $C$ as claimed.

\subsection{Proof of Theorem 3}
Recall that for each dimension $j=1,\ldots, d$, the set $S_j = \{k: \beta_{jk} \neq 0\}$ the set of indices corresponding to non-zero components of the true dimension $\bbeta_j$. Any index set $\mathcal{S} \subset \{1, \ldots, p\}$ such that  $\mathcal{S} \not\supset S_j$ is referred to as an underfitted index set, while any $\mathcal{S} \supsetneq S_j$ other than $S_j$ itself is referred to as an overfitted index set. Correspondingly, we can partition the values of the tuning parameter $\mu_j$ into the underfitted, true, and overfitted ranges respectively,
\[
\Omega_{j-} = \{\mu_j: \hat{S}(\mu_j) \not\supset S_j \},~ \Omega_{0j} = \{\mu_j: \hat{S}(\mu_j)  = S_j \},~ \text{and}~ \Omega_{j+} = \{\mu_j: \hat{S}(\mu_j)  \supsetneq S_j  \}
\] 
where $\hat{S}(\mu_j) = \{k: \hat{\bbeta}^*_{jk} (\mu_j) \neq 0\}$, the set of indices corresponding to the nonzero component of $\hat{\bbeta}^*_j(\mu_j)$, the adaptive Cholesky matrix penalization estimator at the tuning parameter $\mu_j$. 
We will show that, for any $\mu_j$ that cannot identify the true model and the value of $\tau_j$  stated in the Theorem 3, the resulting $\text{PIC}(\mu_j; \tau_j)$ is consistently larger than $\text{PIC}(\mu_{0};\tau_j)$ with $\mu_0 \in \Omega_{0j}$. To simplify the notation, we use $\text{PIC}(\mu_j)$. We will treat two cases of overfitting and underfitting separately. 
\subsubsection*{Overfitted range}
For $\mu_j \in \Omega_{j+}$ (the overfitted range), we have,
\begin{equation}
	\text{PIC}(\mu_j) - \text{PIC}(\mu_0) = \left\|\mathcal{P}\left\{\hat{ \bbeta}_j^*(\mu_j)\right\}-\mathcal{P}(\bar{ \bbeta}_j)\right\|_F^2- \left\|\mathcal{P}\left\{\hat{ \bbeta}_j^*({\mu_0})\right\} -\mathcal{P}(\bar{ \bbeta}_j)\right\|_F^2 + \tau_j\Delta_j
	\label{eq1}
\end{equation}
where $\Delta_j = \norm{\hat{ \bbeta}_j^*(\mu_j)}_0-\norm{\hat{ \bbeta}_j^*(\mu_0)}_0 > 0$. By the triangle inequality, we obtain
\begin{equation}
	\left\|\mathcal{P}\left\{\hat{ \bbeta}_j^*({\mu_0})\right\} -\mathcal{P}(\bar{ \bbeta}_j)\right\|_F \leq \norm{\mathcal{P}\left\{\hat{ \bbeta}_j^*({\mu_0})\right\}-\mathcal{P}( \bbeta_j)}_F + \norm{\mathcal{P}( \bbeta_j)-\mathcal{P}(\bar{ \bbeta})}_F,
	\label{eq:13}
\end{equation} 
For the first term in the right hand side of \eqref{eq:13}, the tuning parameter $\mu_0$ satisfies the condition for the tuning parameter in Theorem 2, so   $\norm{\mathcal{P}\left\{\hat{ \bbeta}_j^*({\mu_0})\right\}-\mathcal{P}(\bbeta_j)}_F = O_p\left[\left\{s\log(p)/(n\lambda)\right\}^{1/2}\right]$. The second term $\norm{\mathcal{P}( \bbeta_j)-\mathcal{P}(\bar{ \bbeta}_j)}_F = O_p(\sqrt{p/n})$. Since $s\log(p) = o(p)$, 
the rate of convergence of the right hand side of  \eqref{eq:13} is dominated by the rate of convergence of the unpenalized estimator; i.e $\norm{\mathcal{P}\left\{\hat{ \bbeta}_j^*({\mu_0})\right\} - \mathcal{P}(\bar{ \bbeta}_j)}_F^2 = O(p/n)$.

Finally, since $\Delta_j>0$ and $\tau_j \stackrel{>}{\sim} p/n$,  the right hand side of  \eqref{eq1} is asymptotically positive, i.e $\text{PIC}(\mu_j) > \text{PIC}(\mu_0)$ for every $\mu_j\in\Omega_+$ when $n \rightarrow \infty$.

\subsubsection*{Underfitted range}
For $\mu_j \in \Omega_{j-}$ (the underfitted range), we want to show
\begin{equation}
	\text{PIC}(\mu_j) - \text{PIC}(\mu_0) = \left\|\mathcal{P}\left\{\hat{ \bbeta}_j^*(\mu_j)\right\}-\mathcal{P}(\bar{ \bbeta}_j)\right\|_F^2- \left\|\mathcal{P}\left\{\hat{ \bbeta}_j^*({\mu_0})\right\} -\mathcal{P}(\bar{ \bbeta}_j)\right\|_F^2+\tau_j\left(\norm{\hat{{\bbeta}}_j^*(\mu_j)}_0-s_j \right) > 0 
	\label{eq:underfitt}
\end{equation}
occurs with probability tending to one as $n \rightarrow \infty$. By the same argument as in the previous section for the overffited range, the second term in the right hand side of  \eqref{eq:underfitt} $\norm{\mathcal{P}\left\{\hat{ \bbeta}_j^*({\mu_0})\right\} - \mathcal{P}(\bar{ \bbeta}_j)}_F^2 = O(p/n)$. For the firm term in \eqref{eq:underfitt}, applying the triangle inequality again, we have 
\begin{equation}
	\norm{\mathcal{P}\left\{\hat{ {\bbeta}}_j^*(\mu_j) \right\} - \mathcal{P}(\bar{ {\bbeta}}_j)}_F \geq \norm{\mathcal{P}\left\{\hat{ {\bbeta}}_j^*(\mu_j) \right\} - \mathcal{P}( {\bbeta}_j)}_F - \norm{\mathcal{P}({ {\bbeta}}_j) - \mathcal{P}(\bar{ {\bbeta}}_j)}_F
	\label{firstterm}
\end{equation}
First, regarding the second term on the right hand side of \eqref{firstterm}, we have $\norm{\mathcal{P}({ {\bbeta}_j}) - \mathcal{P}(\bar{ {\bbeta}}_j)}_F = O(\sqrt{p/n})$. For the first term on the right hand side of \eqref{firstterm}, let $\mathcal{K}_j$ be the index set of underfitted components, i.e for all $k \in \mathcal{K}_j$, we have  $\hat\beta^*_{jk}{(\mu_j)}=0$ while $\beta_{jk} \neq 0$. Hence, all the elements whose at least one of the column and row indices of the estimated projection matrix $\mathcal{P}\left\{\hat{ {\bbeta}_j^*}(\mu_j)\right\}$ are zero.  Therefore, we obtain
$$
\begin{aligned}
	\norm{\mathcal{P}\left\{\hat{ {\bbeta}}_j^*(\mu_j) \right\} - \mathcal{P}( {\bbeta}_j)}_F^2 & \geq 2 \dfrac{\sum_{k \in \mathcal{K}_j}\bbeta_{jk}^2}{\norm{\bbeta_j}_2^2}  - \dfrac{\sum_{k \in \mathcal{K}_j}\bbeta_{jk}^4}{\norm{\bbeta_j}_2^4} - 2 \dfrac{\sum_{k,r \in \mathcal{K}_j}\beta_{jk}^2 \beta_{jr}^2}{\norm{\bbeta_j}_2^4}  \\ & =  2
	\dfrac{\norm{\bm{\beta}_{j\mathcal{K}_j}}_2^2}{\norm{\bbeta_j}_2^2} - \dfrac{\norm{\bm{\beta}_{j\mathcal{K}_j}}_2^4}{\norm{\bbeta_j}_2^4}.
\end{aligned}
$$
Let $\xi_j = \min_{k= 1,\ldots,s_j}\left\{ \dfrac{{\bbeta}_{jk}^2}{ \bbeta_j^\top \bbeta_j} \right\}$, so we have 
$$\vert \mathcal{K}_j \vert \xi_j \leq \frac{ \norm{\bbeta_{j\mathcal{K}_j}}_2^2}{\norm{\bbeta_j}_2^2} < 1,$$ 
where $\vert \mathcal{K}_j \vert $ denotes the cardinality of the set $\mathcal{K}_j$. Since the function $f(x) = x(2-x)$ is monotonic increasing on $[0, 1]$, we obtain
$$
\norm{\mathcal{P}\left\{\hat{ {\bbeta}}_j^*(\mu_j) \right\} - \mathcal{P}( {\bbeta}_j)}_F^2 \geq 2\vert \mathcal{K}_j \vert \xi_j-  \vert \mathcal{K}_j \vert^2 \xi_j^2.
$$
Note that $\norm{\hat{ {\bbeta}}_j^*(\mu_j)}_0 \geq s_j - \vert \mathcal{K}_j \vert $, so when $p/n \rightarrow 0$, equation \eqref{eq:underfitt} is satisfied if for all $\vert \mathcal{K}_j \vert = 1,\ldots, s_j$, we have
\begin{equation}
	2\vert \mathcal{K}_j \vert \xi_j -  \vert \mathcal{K}_j \vert^2 \xi_j^2  - \tau_j\vert\mathcal{K}_j \vert > 0, ~ \text{i.e.} ~  \tau_j < 2\xi_j -  \vert \mathcal{K}_j \vert \xi_j^2 = \xi_j(2- \vert \mathcal{K}_j \vert \xi_j).  
	\label{eq:sj}
\end{equation}
Since $\vert \mathcal{K}_j \vert \xi_j$ is smaller than $1$, equation \eqref{eq:sj} is satisfied if $\tau_j < \xi_j$. In other words, if $\tau_j = o(\xi_j)$, then $\text{PIC}(\mu_j) > \text{PIC}(\mu_0)$ as $n\rightarrow\infty$ as claimed.

\section{A brief review of the Lasso sliced inverse regression estimator}	
In this section, we briefly review the Lasso sliced inverse regression (SIR) estimator and establish the connection between it and the CHOMP estimator.  Assume that a random sample $( \bm{x}_i^\top, y_i),~i=1,\ldots,n$ is generated from  the single index model
$ y_i = f( \bm{x}_i^\top  \bbeta_0,\varepsilon_i),~ i=1,\ldots,n,
$
with the outcome $y_i$, and covariate $ {x}_i$ follows a $p$-dimensional elliptical distribution with location zero and scale matrix $ {\bSigma}$.  Let $\mathcal{X}$ denote the $n\times p$ design matrix. The sliced inverse regression estimate for $ {\bbeta}_0$ is based on the relationship 
\begin{equation}
	{\bSigma} \bbeta_0 \propto  {\boldeta}.
	\label{eq:SIR}
\end{equation}
The covariance matrix $ {\bSigma}$ is estimated by the sample covariance matrix $\hat{ {\bSigma}} = n^{-1} \mathcal{X}^\top \mathcal{X}$. Next, without loss of generality, assume the data $( {\bm{x}}_i, y_i)$ are arranged such that $y_1 \leq y_2 \leq \ldots \leq y_n$. Then the data are divided into $H$ equal-sized slices, denoted by $J_1, \ldots, J_H$ based on the increasing order of $y$. For ease of notation and arguments, assume $n=cH$ with $c>0$. Next, construct a $H \times n $ matrix $ \bm{M} =    \textbf{I}_H \otimes \textbf{1}_c^\top  $, where $ \textbf{I}_H$ denotes the identity matrix of dimension $H$, $ \textbf{1}_c$ denotes the $c \times 1$ vector with all entries being one, and $\otimes$ denotes the outer product. Next, compute the averages of covariates within each slice, $\bar{ {\bm{x}}}_h^\top = c^{-1} \sum_{i=1}^{n}  {\bm{x}}_i^\top1(y_i \in J_h)$, and form a $H \times p$  matrix $\mathcal{X}_H$ with each row being $\bar{ {\bm{x}}}_h^\top,~h=1,\ldots,H$. With this formulation, $\mathcal{X}_H =  \bm{M}\mathcal{X}/c$, so the conditional expectation $ \bm\Lambda = \text{var}\left\{\mathbb{E}(\bX|y)\right\}$ is estimated by
\[
\hat{ \bm\Lambda}={H}^{-1}\sum_{h=1}^{H}  \bar{ \bm{\bm{x}}}_h\bar{ {\bm{x}}}_h^\top = \dfrac{1}{H}\mathcal{X}_H^\top \mathcal{X}_H = \dfrac{1}{nc} \mathcal{X}^\top  \bm{M}^\top \bm{M} \mathcal{X}.
\]
Let $\hat{{\lambda}}$ and $\hat{ {\boldeta}}$ be the largest eigenvalue and its corresponding eigenvector of length one of $\hat{ \bm{\Lambda}}.$ Then 
\[\hat{{\lambda}}\hat{ {\boldeta}} = \hat{ \bm\Lambda}\hat{ {\boldeta}} = \dfrac{1}{nc} \mathcal{X}^\top  \bm{M}^\top \bm{M} \mathcal{X}\hat{ {\boldeta}}.
\]	
Let $\tilde{ \bm{y}} = (c\hat{\lambda})^{-1} \bm{M}^\top \bm{M} \mathcal{X}\hat{ {\boldeta}}$, then we have $ {\hat{\boldeta}} = n^{-1} \mathcal{X}^\top \tilde{ \bm{y}}$. Therefore, the estimated version of equation \eqref{eq:SIR} can be written as $\mathcal{X}^\top \mathcal{X} {\bbeta}_0 \propto  \mathcal{X}^\top \tilde{\bm{y}}$ and the Lasso SIR estimate is defined as
\begin{equation*}
	\hat{ {\bbeta}}^{\text{L}}=\arg\min_{ {\bbeta}}\frac{1}{2 n}\left\|\tilde{\bm{y}}-\mathcal{X} {\bbeta}\right\|_{2}^{2}+\mu\|{\bbeta}\|_{1},
\end{equation*}
with $\mu$ being an appropriate tuning parameter. For the multiple index model $y_i = f( {x}_i^\top  \bbeta_1, \ldots,  {x}_i^\top  \bbeta_d, \varepsilon_i)$, the Lasso SIR estimator for each dimension is defined as 
\begin{equation}
	\hat{ {\bbeta}}_j^{\text{L}} = \arg\min_{ {\bbeta}_j} \frac{1}{2n}\left\|\widetilde{{y}}_{j}-\mathcal{X} {\bbeta}_j\right\|_{2}^{2}+\mu_{j}\|{\bbeta_j}\|_{1},~j=1,\ldots,d,
	\label{eq: SIRLasso}
\end{equation}
where $\widetilde{ \bm{y}}_j = (c\hat{\lambda}_j)^{-1} {M}^\top {M} \mathcal{X}\hat{ {\boldeta}}_j$, with $\hat{\lambda}_j$ and $\hat{ {\boldeta}}_j$ being the $j^{th}$ largest eigenvalue and its corresponding eigenvector of $\hat{ \bm{\Lambda}}$, and the $\mu_j$ are tuning parameters. 

Next, we show that the Lasso SIR has the same estimating equation as the CHOMP estimate. From the definition \eqref{eq: SIRLasso} and the first order condition, each component of the Lasso SIR $\hat{\bbeta}_{j}^L = (\hat{\bbeta}_{j1}^L, \ldots, \hat{\bbeta}_{jp}^L)^\top$ satisfies
$$
n^{-1} \bm{x}_k^\top (\tilde{\bm{y}} - \mathcal{X}\hat{\bbeta}_j) + \mu_j b^L_{jk} = 0,
$$
where $b^L_{jk} = \text{sign}(\hat{\bbeta}_{jk})$ if $\hat{\bbeta}_{jk} \neq 0$ and $b^L_{jk} \in [-1, 1]$ otherwise. Also, $n^{-1} \bm{x}_k^\top \tilde{\bm{y}} = \tilde{\boldeta}_{jk}$, and $n^{-1} \bm{x}_k^\top \mathcal{X} = \hat{\bSigma}_k$, the $k^{th}$ row of the sample covariance matrix $\bSigma$. Hence, for any tuning parameter $\mu_j$, the estimating equation of the  Lasso SIR is 
$$
-\hat\bSigma_k \hat{\bbeta}_j + \hat\boldeta_{jk} + \mu_j b_{jk}^L = 0
$$
exactly the same as the estimating equation of the CHOMP estimator shown in the paper. As a result, it is not surprising that the CHOMP and the Lasso SIR estimator require the same theoretical value of tuning parameters to ensure estimation consistency and share the same convergence rate. 

Similar to any regularization method, the performance of the Lasso SIR depends critically on the choice of the tuning parameters $\mu_j$. In their simulation study, \cite{Linsparse} implemented \eqref{eq: SIRLasso} as a lasso problem with design matrix $\mathcal{X}$ and outcome $\tilde{ {y}_j}$ and used ten-fold cross-validation  to choose the tuning parameters $\mu_j$. We show via a small simulation below that the Lasso SIR estimator with this choice of tuning parameters has performance close to the lasso sliced inverse regression estimator where tuning parameters are chosen optimally. This finding justifies our comparison of the Lasso SIR estimator estimator with tuning parameter selected via ten-fold cross-validation with other estimators in the simulation study of the main paper. 

For the simulation study, we generate independent and identically distributed data $( \bm{x}_i^\top,y_i$)  as in the single index model simulation in Section 5.1 of the main paper with $s=5$ and $n=500$. The number of slices is fixed at $H=20$ and the number of indices $d=1$ is assumed to be known. We compare the average estimation error across $1000$ samples of the Lasso SIR estimator under two methods for choosing the parameter. For the first method, the tuning parameter is chosen through ten-fold cross-validation. For the second method, the tuning parameter is chosen to minimize the actual estimation error; this choice of tuning parameter is referred to as the optimal tuning parameter. For any tuning parameter $\mu$, the estimation error is defined as $\textrm{Error}=\norm{\mathcal{P}(\hat{ {\bbeta}}_\mu^{\text{L}}) - \mathcal{P}({ {\bbeta}_0})}_F^2$, the squared Frobenius norm of the difference between the estimated projection matrix and the true projection matrix. Note that the optimal tuning parameter is not available in practice, because it requires knowledge of the true vector $ {\bbeta}_0$. 

\begin{table}[ht]
	\centering
	\caption{Estimation error of the Lasso SIR estimator with tuning parameter selected via ten-fold cross-validation and with optimal tuning parameter. Standard errors are in parentheses}
	\begin{tabular}{cccc}
		$p$ & $ {\bbeta}_0$ & Cross-validation & Optimal \\[3pt]
		\hline
		40 & Large & 0.28 (0.06) & 0.27 (0.06) \\
		& Small & 0.43 (0.10) & 0.41 (0.09) \\[8pt]
		
		100 & Large & 0.36 (0.06) & 0.34 (0.06) \\
		& Small & 0.55 (0.10) & 0.54 (0.10)  \\
		\hline
	\end{tabular}
	
	\label{tab:sirlassocompare}
\end{table}
It can be seen that the Lasso SIR estimator with tuning parameter selected via cross-validation gives very similar performance to the same estimator with optimal tuning parameter, where the difference in estimation error is negligible. This result is surprising given the pseudo response $\tilde{y}$ does not contain independent components, so investigating why cross-validation still works for the lasso sliced inverse regression estimation can be a topic for future research. 

\section{CHOMP for SIR in high dimensional settings}
\label{sir-highdim}
\subsection{Method}
In this section, we demonstrate how the CHOMP technique can be extended to sufficient dimension methods such as the SIR in high dimensional settings. In such scenario, one particular challenge of implementing the CHOMP and the adaptive CHOMP estimators is to find a good estimator for the Cholesky factor $ \bm{L}$ of the population covariance matrix $ \bSigma$ and its inverse.  While this is hard in general, we can estimate $\bm{L}$ efficiently when the population covariance matrix has some special structure.

In this section, we consider a regression  setting where the covariates have a natural order (for example when they are collected over time) and the population covariance (and correlation) matrix are banded, i.e $\sigma_{jk} = 0$ if $\vert j-k \vert > K$ with $K$ known. Such covariance structure has been considered extensively in the literature of high-dimensional covariance estimation, see for example \cite{pourahmadi2013high} and \citet{khare2019scalable}. In this case, let $ {\bSigma} =  \bm{CDC}^\top$ be the modified Cholesky decompositon of $ {\bSigma}$ such that $ \bm{D}$ is a diagonal matrix and $ \bm{C} = (c_{jk})$ is a lower triangular matrix with $c_{jj} = 1$ and $c_{jk} = 0$ if $\vert j-k \vert >K$. As suggested by \cite{rothman2010new}, the off-diagonal elements of $ \bm{C}$ and the diagonal elements of $\bm{D}$ can be estimated sequentially by fitting a sequence of linear regressions. Let $\bm{x}^{(j)}, j = 1,\ldots, p$ denote the $j$th column of the design matrix $\mathcal{X}$. For the first variable, set $\bm{e}_1 = \bm{x}^{(1)}$. For $j=2,\ldots,p$, let $\bm{c}_j^{(k)} = (c_{j, j-k}, \ldots, c_{j,j-1})^\top$ and $\bm{Z}_j^{(k)} = (\bm{e}_{j-k}, \ldots, \bm{e}_{j-1})$, where the index $j-k$ is understood to mean ${\max}(1, j-k)$, then we compute sequentially
$$
\bm{\hat{c}}_j^{(k)} = \arg\min_{\bm{c}_j^{(k)}} \norm{\bm{x}^{(j)} - \bm{Z}_j^{(k)}\bm{c}_j^{(k)}}_2^2, \bm{e}_j = \bm{x}^{(j)} - \bm{Z}_j^{(k)}\hat{\bm{c}}_j^{(k)}.
$$
Finally the diagonal elements of $\bm{D}$ are estimated as $\hat{d}_{jj} = n^{-1}\norm{\bm{e}_j}_2^2$, and the Cholesky factor $ \bm{L}$ is estimated by $\hat{ \bm{L}} = \widehat{ \bm{C}}\widehat{ \bm{D}}^{1/2}$, where $\widehat{ \bm{D}} = \text{diag}(\hat{d}_jj)$ and $\widehat{ \bm{C}} = (\hat{c}_{jk}), ~ j,k =1,\ldots,p$. 

Let $\hat{\bm\kappa}_j$ be calculated such as $\hat{\bm{L}}\hat{\bkappa}_j = \hat{\bm\eta}_j$, where $\bm{\eta}_j$ is calculated in the same way as outlined in Section 3.1 of the main paper. With the regression-based estimated Cholesky factor $\hat{\bm{L}}$, the CHOMP and adaptive CHOMP estimator for SIR are defined respectively as
\begin{equation}
	\hat{ \bbeta}_j =  \arg\min_{ \bbeta_j} \dfrac{1}{2} \norm{\hat{ \bm{L}}^\top {\bbeta}_j - \hat{ {\bkappa}}_j}_2^2 + \mu_j \norm{ {\bbeta}_j}_1, ~ j=1,\ldots,d, 
	\label{eq:Cholesky matrix  penalization_2}
\end{equation}
and 
\begin{equation}
	\hat{ \bbeta}_j^* = \arg\min_{ \bbeta_j} \dfrac{1}{2} \norm{\hat{ \bm{L}}^\top {\bbeta}_j - \hat{ {\bkappa}}_j}_2^2 + \mu_j \sum_{k=1}^{p} \omega_{jk} \vert \beta_{jk} \vert , ~ j=1,\ldots,d. 
	\label{eq:adaCholesky matrix  penalization_2},
\end{equation}
where $\omega_{jk}$ are $\omega_{jk} = \vert \bar{\beta}_{jk} \vert^{-\gamma}$, with $\bar{\beta}_{jk}$ the $k${th} component of an initial consistent estimate $\bar{ {\bbeta}}_j$ and $\gamma$ a positive constant. Inhigh dimensional settings, the unpenalized sliced inverse regression estimator is not consistent \citep{lin2018consistency}. Hence, for each dimension, we use the Lasso SIR estimator as the initial consistent estimator $\bar{\bm\beta}_j$ for computing the adaptive weight. Furthermore, to adjust for the convergence rate of the Lasso SIR estimator, we select the tuning parameters for the CHOMP and adaptive CHOMP from minimizing the following projection information criterion
\begin{equation*}
	\textsc{PIC}(\mu_j; \tau_j) = \begin{cases}
		\norm{\mathcal{P}\left\{\hat{ \bbeta}({\mu_j})\right\}-\mathcal{P}(\bar{ \bbeta}_j)}_F^2 + \dfrac{2}{p} \norm{\hat{ \bbeta}_j({\mu_j})}_0,& ~ \text{if}~ \hat{ \bbeta}({\mu_j}) \neq  {0}  \label{first} \\
		\infty, & ~ \text{if} ~ \hat{ \bbeta}_j({\mu_j}) =  {0}.
	\end{cases}
\end{equation*}
Estimation and selection consistency of the CHOMP-based estimators and of the PIC in high dimensional settings where the Cholesky factors are estimated based on regression will be topics of future research. Below, we will present a simulation study to demonstrate the empirical performance of this approach.

\subsection{Simulation}
For the simulation, we generate data from the model (I) as in section 5.1 of the main paper with the correlation matrix $\tilde{ \bm{\Omega}}$ having off-diagonal elements $(\tilde{\omega})_{jk} = 1- K^{-1}\vert j-k \vert$  if $\vert j-k \vert \leq K$ and 0 otherwise. We consider two values for $K$, namely $K \in \{3,5\}$. The sample size is fixed at $n=1000$ and the number of covariates varies over $p \in \{500, 1000, 1500\}.$ We compute the CHOMP, adaptive CHOMP estimator with $\gamma=1$ and $\gamma =2$, and the Lasso SIR estimators; then we compare them using the same metric as in Section 5.1 of the main paper.

\begin{table}[ht]
	\centering
	\caption{Performance of the estimators in the simulation of single index model in high dimensional settings where the covariance of the covariates are banded. Standard errors are in parentheses. The lowest estimation error is highlighted for each setting.}
	\resizebox{\linewidth}{!}{\begin{tabular}{lll ccccc}
			\toprule[1.5pt]
			$p$ & $K$ & Metric & CHOMP & \multicolumn{2}{c}{Adaptive CHOMP} & Lasso SIR & Mlasso \\
			&&&& $\gamma=1$& $\gamma=2$ & & \\ 
			\hline \addlinespace
			500 & 3 & Error & 0.89 (0.22) & 0.45 (0.23) & \textbf{0.39} (0.17) & 0.44 (0.15) & 0.99 (0.22) \\ 
			&  & FPR & 0.00 (0.00) & 0.00 (0.00) & 0.00 (0.00) & 0.10 (0.06) & 0.01 (0.02) \\ 
			&  & FNR & 0.29 (0.18) & 0.05 (0.12) & 0.02 (0.07) & 0.02 (0.05) & 0.40 (0.23) \\ 
			& 5 & Error & 0.99 (0.21) & 0.50 (0.28) & \textbf{0.43} (0.22) & 0.49 (0.19) & 1.06 (0.23) \\ 
			&  & FPR & 0.00 (0.00) & 0.00 (0.01) & 0.00 (0.01) & 0.10 (0.06) & 0.01 (0.01) \\ 
			&  & FNR & 0.42 (0.24) & 0.12 (0.24) & 0.08 (0.20) & 0.03 (0.09) & 0.51 (0.26) \\[3pt] 
			
			1000 & 3 & Error & 1.06 (0.17) & 0.53 (0.27) & \textbf{0.43} (0.19) & 0.51 (0.17) & 1.01 (0.19) \\ 
			&  & FPR & 0.00 (0.00) & 0.00 (0.00) & 0.00 (0.00) & 0.07 (0.04) & 0.01 (0.01) \\ 
			&  & FNR & 0.48 (0.21) & 0.09 (0.17) & 0.04 (0.09) & 0.03 (0.08) & 0.39 (0.22) \\ 
			& 5 & Error & 1.07 (0.18) & 0.61 (0.31) & \textbf{0.52} (0.25) & 0.57 (0.21) & 1.06 (0.22) \\ 
			&  & FPR & 0.00 (0.00) & 0.00 (0.00) & 0.00 (0.01) & 0.07 (0.04) & 0.00 (0.01) \\ 
			&  & FNR & 0.53 (0.24) & 0.18 (0.27) & 0.09 (0.17) & 0.05 (0.11) & 0.52 (0.26) \\[3pt] 
			
			1500 & 3 & Error & 1.05 (0.18) & 0.77 (0.29) & \textbf{0.57} (0.25) & 0.61 (0.24) & 0.97 (0.25) \\ 
			&  & FPR & 0.00 (0.00) & 0.00 (0.00) & 0.00 (0.00) & 0.04 (0.03) & 0.00 (0.00) \\ 
			&  & FNR & 0.52 (0.24) & 0.29 (0.27) & 0.13 (0.18) & 0.11 (0.17) & 0.40 (0.24) \\ 
			& 5 & Error & 1.11 (0.16) & 0.84 (0.30) & \textbf{0.62} (0.25) & 0.64 (0.23) & 1.03 (0.23) \\ 
			&  & FPR & 0.00 (0.00) & 0.00 (0.00) & 0.00 (0.00) & 0.04 (0.03) & 0.00 (0.00) \\ 
			&  & FNR & 0.59 (0.23) & 0.36 (0.29) & 0.15 (0.15) & 0.13 (0.15) & 0.48 (0.27) \\
			\bottomrule[1.5pt]
	\end{tabular}}
	\vspace{0.2cm}
	\label{tab:high_dimension}
\end{table}
\FloatBarrier
Table \ref{tab:high_dimension} demonstrates that the adaptive CHOMP estimator with $\gamma = 2$ has the best performance in the considered settings. The Matrix Lasso and the CHOMP estimate tend to have approximately the same estimation error, which is usually higher than both the adaptive CHOMP and the Lasso SIR due to higher false negative rates. Compared to the Lasso SIR estimator, both the adaptive CHOMP estimators with $\gamma=1$ and $\gamma=2$ tend to reduce the false positive rates. However, the adaptive CHOMP with $\gamma=1$ tends to underfit by having medium false negative rates (as seen in $p=1500$). On the other hand, the adaptive CHOMP estimator with $\gamma=2$ does not increase the false negative rate much and hence has the lowest estimation error. 

\newpage
\section{Additional Simulation Results}
\begin{table}[ht]
	\centering
	\caption{Performance of the estimators in the single index model simulation in section 5.1 of the main paper with the correlation matrix $\tilde{ \bm{\Omega}}$ having homogeneous structure. Standard errors are included in parentheses. The lowest estimation error in each setting is highlighted.}
	\resizebox{.9\linewidth}{!}{\begin{tabular}{lllccccc}
			\toprule[1.5pt]
			Model & $p$ & Metric &CHOMP &
			\multicolumn{2}{c}{Adaptive CHOMP} & Lasso SIR & Mlasso \\
			
			&&&& $\gamma=1$& $\gamma=2$ &&  \\ 
			\hline \addlinespace
			(I) & 100 & Error & 0.59 (0.31) & \textbf{0.18} (0.18) & 0.19 (0.18) & 0.24 (0.15) & 0.83 (0.28) \\ 
			&  & FPR & 0.00 (0.01) & 0.01 (0.04) & 0.02 (0.05) & 0.16 (0.08) & 0.01 (0.02) \\ 
			&  & FNR & 0.16 (0.24) & 0.01 (0.10) & 0.01 (0.08) & 0.01 (0.06) & 0.33 (0.28) \\ 
			& 200 & Error & 1.08 (0.29) & \textbf{0.26} (0.28) & 0.33 (0.27) & 0.28 (0.22) & 0.98 (0.28) \\ 
			&  & FPR & 0.00 (0.00) & 0.02 (0.04) & 0.05 (0.07) & 0.10 (0.05) & 0.00 (0.01) \\ 
			&  & FNR & 0.62 (0.30) & 0.04 (0.19) & 0.02 (0.13) & 0.02 (0.13) & 0.49 (0.29) \\[3pt] 
			
			(II) & 100 & Error & 0.50 (0.28) & \textbf{0.07} (0.14) & \textbf{0.07} (0.14) & 0.09 (0.13) & 0.78 (0.31) \\ 
			&  & FPR & 0.00 (0.00) & 0.00 (0.02) & 0.00 (0.03) & 0.18 (0.08) & 0.01 (0.02) \\ 
			&  & FNR & 0.12 (0.17) & 0.01 (0.08) & 0.01 (0.07) & 0.01 (0.07) & 0.31 (0.28) \\ 
			& 200 & Error & 0.84 (0.30) & \textbf{0.10} (0.24) & 0.11 (0.24) & 0.12 (0.21) & 0.95 (0.31) \\ 
			&  & FPR & 0.00 (0.00) & 0.01 (0.03) & 0.01 (0.05) & 0.11 (0.05) & 0.00 (0.01) \\ 
			&  & FNR & 0.35 (0.30) & 0.03 (0.15) & 0.02 (0.11) & 0.01 (0.10) & 0.48 (0.30) \\[3pt]
			
			(III) & 100 & Error & 0.51 (0.28) & \textbf{0.08} (0.14) & \textbf{0.08} (0.14) & 0.11 (0.13) & 0.79 (0.30) \\ 
			&  & FPR & 0.00 (0.01) & 0.00 (0.03) & 0.00 (0.03) & 0.17 (0.08) & 0.01 (0.02) \\ 
			&  & FNR & 0.12 (0.18) & 0.00 (0.06) & 0.00 (0.05) & 0.01 (0.06) & 0.31 (0.27) \\ 
			
			& 200 & Error & 0.86 (0.31) & \textbf{0.12} (0.24) & \textbf{0.12} (0.24) & 0.14 (0.20) & 0.95 (0.30) \\ 
			&  & FPR & 0.00 (0.00) & 0.01 (0.04) & 0.01 (0.05) & 0.10 (0.05) & 0.00 (0.01) \\ 
			&  & FNR & 0.37 (0.31) & 0.03 (0.14) & 0.02 (0.12) & 0.02 (0.11) & 0.47 (0.30) \\ 	
			\bottomrule[1.5pt]
	\end{tabular}}
	\vspace{0.2cm}
	\label{tab:tab3}
\end{table}
\FloatBarrier


\end{document}